\documentclass[aps,prb,reprint,superscriptaddress,
raggedbottom]{revtex4-2}
\usepackage{graphicx}
\usepackage{amsmath}
\usepackage{amssymb}
\usepackage{xspace}
\usepackage{xcolor} 
\usepackage{upgreek}

\usepackage[utf8]{inputenc}

\newcommand{\beq}{\begin{equation}}
\newcommand{\eeq}{\end{equation}}

\usepackage{pgffor} % To support \foreach coommand in macros
\DeclareRobustCommand{\chem}[1]{%
\ensuremath{\foreach \x/\y in {#1} {{\mathrm{\x}_{\y}}}}\xspace}

\newcommand{\usr}[2]{\ensuremath{#1_{\mathrm{#2}}}\xspace} % 

\newcommand{\kmo}{\chem{K/0.3,Mo/,O/3}}

\newcommand{\unit}[1]{\ensuremath{\,{\mathrm{#1}}}\xspace}
\newcommand{\unitx}[1]{\ensuremath{{\mathrm{#1}}}\xspace}
\newcommand{\pcm}{\ensuremath{\,\mathrm{cm}^{-1}}\xspace}
\newcommand{\mum}{\unit{\mu m}}
\newcommand{\mumx}{\unitx{\mu m}}

\newcommand{\psec}{\unit{ps}}

\newcommand{\THz}{\unit{THz}}
\newcommand{\Kel}{\ensuremath{\,\mathrm{K}}}

\newcommand{\tisa}{Ti:Al$_2$O$_3$\xspace}

\newcommand{\rarr}{\ensuremath{\rightarrow}\xspace}
\newcommand{\simx}{{\sim}}

\newcommand{\kF}{\usr{k}{F}\xspace}
\newcommand{\epsr}[1]{\usr{\varepsilon}{#1}\xspace}
\newcommand{\nuzn}{\ensuremath{\nu_{0n}}\xspace}
\newcommand{\nuznA}{\ensuremath{\nu_{0n}^{(A)}}\xspace}
\newcommand{\nuznP}{\ensuremath{\nu_{0n}^{(P)}}\xspace}
\newcommand{\nuzA}{\ensuremath{\nu_{01}^{(A)}}\xspace}
\newcommand{\nuzP}{\ensuremath{\nu_{01}^{(P)}}\xspace}
\newcommand{\wzn}{\ensuremath{\omega_{0n}}\xspace}
\newcommand{\Wzn}{\ensuremath{\Omega_{0n}}\xspace} % Bare mode frequencies
\newcommand{\Gamn}{\ensuremath{\Gamma_{n}}\xspace}
\newcommand{\phin}{\ensuremath{\varphi_{0n}}\xspace}
\newcommand{\Wpin}{\usr{\Omega}{p}\xspace}

\newcommand{\ffm}{Physikalisches Institut, J. W. Goethe-Universität, D-60438 Frankfurt am Main, Germany} 
\newcommand{\jgu}{Institute of Physics, Johannes Gutenberg-University Mainz, D-55128 Mainz, Germany}
\begin{document}

\title{Combined investigation of collective amplitude and phase modes in a quasi-one-dimensional charge-density-wave system over a wide spectral range}

\author{Konstantin Warawa}
\affiliation{\ffm}
\author{Nicolas Christophel}
\affiliation{\ffm}
\author{Sergei Sobolev}
\affiliation{\jgu}
\author{Jure Demsar}
\affiliation{\jgu}
\author{Hartmut G. Roskos}
\affiliation{\ffm}
\author{Mark D. Thomson}
\affiliation{\ffm}
%\date{January 2023}

\begin{abstract}
We investigate experimentally both the amplitude and phase channels of the collective modes in the quasi-1D charge-density-wave (CDW) system, \kmo, by combining (i) optical impulsive-Raman pump-probe and (ii) terahertz time-domain spectroscopy (THz-TDS), with high resolution and a detailed analysis of the full complex-valued spectra in both cases.
This allows an unequivocal assignment of the observed bands to CDW modes across the THz range up to $9\THz$.
We revise and extend a time-dependent Ginzburg-Landau model to account for the observed temperature dependence of the modes, where the combination of both amplitude and phase modes allows one to robustly determine the bare-phonon and electron-phonon coupling parameters.  
While the coupling is indeed strongest for the lowest-energy phonon, dropping sharply for the immediately subsequent phonons, it grows back significantly for the higher-energy phonons, demonstrating their important role in driving the CDW formation.
We also include a reassessment of our previous analysis of the lowest-lying phase modes, whereby assuming weaker electronic damping for the phase channel results in a qualitative picture more consistent with quantum-mechanical treatments of the collective modes, with a strongly coupled amplitudon and phason as the lowest modes.
\end{abstract}  

\maketitle

\section{Introduction}

Charge density waves (CDWs) constitute an important example of symmetry-broken ground states, arising in low-dimensional conductors and typically driven by electron-phonon (e-ph) coupling,
manifesting as an electron-density modulation and periodic lattice distortion (PLD) with wavevectors $q=2\kF$.  Their study continues to take on new relevance, especially as they can appear as co-existing/competing phases in complex solids, e.g. %
unconventional superconductors \cite{chang_direct_2012,lee19,ortiz20, chen21,yu21,liu_tas2_21}, nematic compounds \cite{sato_thermodynamic_2017, yao22} and Weyl semimetals \cite{gooth19}.
%including cuprates \cite{chang_direct_2012}, 
%122-pnictides \cite{lee19}, 
%kagome metals (?) \cite{ortiz20, chen21, yu21}, 
%TMDs \cite{liu_tas2_21}. 
%Nematicity \cite{sato_thermodynamic_2017, yao22}. 
%Weyl semimetal ("axion insulator") \cite{gooth19}. 
%Recent ultrafast studies on those systems (non-equilibrium studies on collective modes (BNA) \cite{pokharel22} and band structure (trARPES on TSI) \cite{crepaldi22}.
Here the low-energy excitations offer an important spectroscopic probe, for both ground-state and non-equilibrium studies \cite{schaefer10,yusupov10,hinton13,thomson17,pokharel22,crepaldi22}. 
In addition to the single-particle gap, corresponding to excitation of electron-hole pairs from the CDW condensate into the adjacent conduction band (and typically lying in the mid-infrared \cite{degiorgi94}),
coupling between phonons and the electronic modulation at $q=2\kF$ gives rise to collective modes at lower energies  -- typically in the terahertz (THz) range -- which serve as a sensitive probe of the CDW physics.
While the PLD alone may lead to zone-folding (and hence allow the appearance of conventional phonons at new energies below the CDW transition temperature $T_c$), the CDW collective modes arise specifically due to e-ph coupling and exhibit physical properties  comprising both the underlying bare phonons and coupled electronic wave, the latter characterized by a complex-valued electronic order parameter (EOP, $\Delta$).
These excitations manifest as both amplitude- and phase-modes (AMs, PMs), which are respectively Raman- and 
infrared-active in centrosymmetric materials. %
While the bare phonons may have a vanishing dipole response vs. their lattice displacements in the normal phase (and hence be only Raman-active), the PMs possess IR-activity as an electromagnetic field can drive them via the polarization of the electron density modulation \cite{rice76,rice78}.
Nevertheless, a reliable assignment of phonon-like bands appearing below $T_c$ to CDW modes is affected by the fact that in the quasi-1D systems, one also has a transition from a normal-phase metal to a semiconducting CDW phase, such that conventional IR-active phonons can also emerge below $T_c$ due to the lifting of screening in the normal phase, and any temperature dependence could, in principle, be due to interaction with the ($T$-dependent) free carriers \cite{rice1979,mihaly1989}.

In order to reliably assign CDW collective modes, a rigorous approach is to demonstrate the simultaneous appearance of both AMs and PMs (in their respective spectroscopies), and ideally also account for their $T$-dependence with an applicable physical model.  
This is the subject of the present report, applied to the well-established quasi-1D CDW system \kmo, using both impulsive-Raman pump-probe spectroscopy and THz-TDS to characterize the AMs and PMs, respectively, with both high spectral resolution and coverage, resolving modes up to $9\THz$ ($\simx 300\pcm$).  
This study extends our previous reports \cite{schaefer10,schaefer14,thomson17}, which were limited to the lowest-frequency modes (${<}3\THz$), 
and provides a comprehensive analysis beyond those in other earlier studies of the Raman-active \cite{travaglini_raman83,sagar_raman08} and IR-active modes \cite{travaglini_ir84,ng86,degiorgi91,beyer2012} in \kmo.

As previously, we employ a phenomenological time-dependent Ginzburg-Landau (TDGL) model, which we now apply to account for the full set of modes and their $T$-dependence, yielding estimates of the e-ph coupling for each bare mode contributing to the manifold.
An important outcome of this study is that while the lowest phonon indeed has the strongest electronic coupling (akin to certain notions in the literature that only a single phonon is involved in forming the CDW \cite{sagar_raman08}), the coupling for the higher-lying phonons first weakens abruptly, but then \textit{increases} with phonon energy, demonstrating their importance for driving the CDW state formation. 

Moreover, the new results for the PM spectra lead us to a significant revision of both the lowest fitted experimental mode and parameters in the TDGL model.
Our previous analysis \cite{thomson17} of the low-lying PMs was based on the differential reflectivity changes in
optical-pump THz-probe experiments, to follow the time evolution of the non-equilibrium response.  Here a strong spectral feature at about $\nu\simx 1.75\THz$ led us to fit the data assuming a PM in this region, which allowed a detailed quantitative fit of both real and imaginary parts of the differential spectra.
The presence of a PM at this location was indeed predicted from the  TDGL model used, assuming strong damping for both the amplitude and phase components of the EOP (discussed in Sec.~\ref{sec:modes}).
One main motivation of the current work was to scrutinize this assignment with high-resolution ground-state THz reflectivity measurements.  Here we find that while such a strong feature is present in the ground-state reflectivity spectrum at this frequency, this can be reproduced by a band model where no PM is present in this vicinity, due to the strong non-local behavior in how modes affect the reflectivity spectrum.
As presented below, this leads us to review the TDGL model, assuming  a significantly weaker damping ($\gamma_2$) for the electronic phase motion, which then predicts that the lowest AM does not have a closely lying PM.  Moreover, this yields predictions for a ``phason'' (i.e. the Goldstone mode, shifted slightly from zero-frequency due to impurity pinning) much more consistent with early experiments \cite{mihaly1989, degiorgi91}, and a qualitative AM/PM arrangement more consistent with quantum-mechanical models \cite{rice76,rice78}.

\section{Experimental details}\label{sec:details}

All experiments were performed on single crystals of \kmo in the incommensurate CDW phase below $T_c=183\Kel$, using complementary time-domain spectroscopy techniques, with radiation polarized along the $b$-axis of the crystal.
The coherent detection of (Raman-active) AMs was realized in all-optical reflective pump-probe experiments, where 40-fs pulses from a 250-kHz \tisa amplifier laser at 800~nm wavelength were used for both optical pump and probe pulses, as described previously \cite{schaefer10}.

To investigate the (IR-active) PMs, we utilized broadband THz-TDS, based on a 1-kHz \tisa{} amplifier laser, using a two-color air plasma for the source
\cite{roskos2007,blank2013,thomson18,thomson23}
and electro-optic sampling (EOS) with 30-fs optical detection pulses, covering a spectral range from $\simx$0.5-7~THz (see Sec.~\ref{sec:thz}, Fig.~\ref{fig:tds}, which includes a schematic of the setup). 
We used a 100-\mumx-thick $\langle 110 \rangle$-cut GaP crystal as the EOS sensor, supported by a 3-mm-thick $\langle 100 \rangle$-cut GaP substrate, to delay signal reflections and provide a time window ${>}40\psec$ for the main THz pulse, and hence an achievable spectral resolution of ${<}25$~GHz.
Additional measurements with air-biased coherent detection (ABCD \cite{karpowicz08}) were employed at $T=20\Kel$ to reach higher THz frequencies (although the signal-to-noise ratio was superior for EOS detection used for the majority of experiments).
The THz beam path comprised four off-axis paraboloidal mirrors for imaging the beam at the sample and detection focal planes.  In order to adapt this transmission-geometry setup for reflectivity measurements, we employed a Au-coated hyperboloidal mirror (of in-house construction) to divert the beam focus onto the sample (angle of incidence $28^\circ$) in a LHe cryostat (equipped with a 50-\mumx-thick polypropylene window), which preserved the subsequent beam propagation to the detector.
Multiple optical guide beams and a camera were used to aid alignment of the sample.  
A linear THz polarizer (vendor: Tydex) was placed in the beam before focusing on the sample to ensure p-polarized THz light along the $b$-axis of the \kmo sample, while the whole setup was purged with dry air in order to suppress the water vapor response in this THz range. 

\section{Amplitude modes: Impulsive Raman scattering}\label{sec:opop}
\begin{figure*}
\centering
\includegraphics[width=\textwidth]{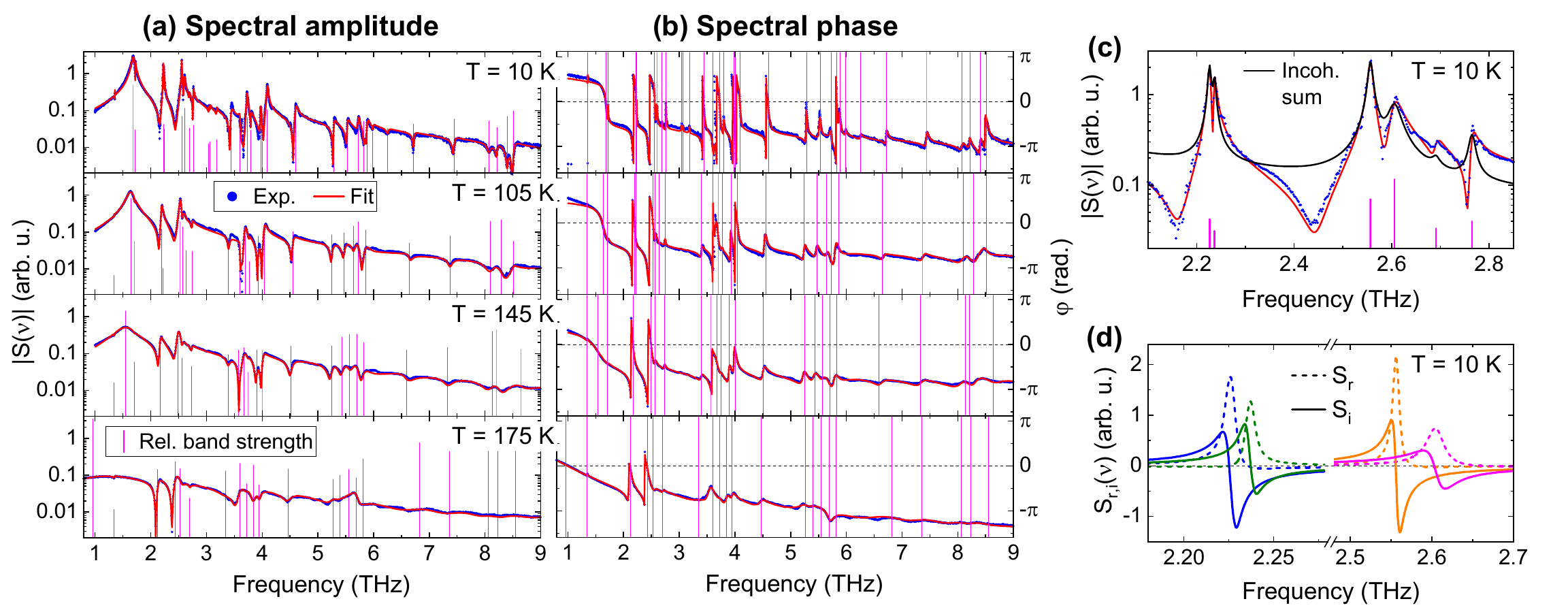}
\caption{
AM spectra from all-optical pump-probe (impulsive Raman) experiments for selected temperatures: (a) Spectral amplitude $|S(\nu)|$ and (b) phase $\varphi(\nu)=\arg{S(\nu)}$, including fit results (red curves). Vertical magenta lines denote fitted mode frequencies (scaled by relative band strength in (a)). %
(c) Magnified range of spectrum in (a) for $T=10\Kel$, including also the 
``incoherent'' spectrum $\usr{S}{Ram}(\nu)$ (black curve) obtained by summing band intensities.
(d) Real ($S_r$, dashed) and imaginary ($S_i$, solid) parts of fitted band spectra for selected modes.
} \label{fig:opop}
\end{figure*}

We begin by presenting the new AM analysis results from previously published, all-optical pump-probe reflectivity experiments \cite{schaefer10,schaefer14}, which allow coherent detection of AMs via their impulsive excitation and subsequent modulation of the optical probe reflectivity (the term ``impulsive'' here implying the general case, covering both impulsive and displacive limits \cite{stevens2002}). 
As mentioned in the Introduction, our ability to perform a new, comprehensive analysis of the mode spectra, i.e. including higher-frequency/weak modes, relies on globally fitting the complex Fourier spectra, as opposed to approaches such as sequential fitting of bands in sub-ranges about their center frequencies.
A summary of our method is given in Appendix~\ref{app:opop}.  
Notable aspects include: (i) One can resolve and characterize (weak) bands even in the presence of significant overlap, (ii) the initial phases $\phin$ of each mode are inherently included via the spectral phase, and (iii) the regression delivers the complex mode amplitudes $A_n$ in each iteration, such that only the mode frequencies ($\wzn$) and bandwidths ($\Gamn$) must be searched, greatly improving the reliability and speed of the algorithm.
Superposed on the time-domain signals $S(t)=\Delta R(t)/R_0$ are non-oscillatory components $\usr{S}{el}(t)$ due to predominantly electronic responses to the excitation. %
While one can incorporate these in the complex spectral analysis (assuming exponential decay kinetics, these manifest as zero-frequency, i.e. Drude-like, bands), we found that the conventional approach of first fitting and subtracting these contributions in the time domain \cite{schaefer10} is advantageous for the subsequent mode analysis, as one must ensure that any (either broadband or low-frequency) spectral background is effectively suppressed to allow fitting of the weak/broad modes, including careful treatment of the initial signal around $t=0$.

Examples of the time-domain signals and the multi-exponential fitting analysis for $\usr{S}{el}$ are given in the Supplementary.
The spectra of the residual oscillatory signals are shown in Fig.~\ref{fig:opop} for selected temperatures below $T_c$, in terms of both (a) the Fourier spectral amplitude $|S(\nu)|$ and (b) phase $\varphi(\nu)$.
(We will explicitly use the terms ``spectral amplitude'' and ``spectral phase'' when discussing the experimental data, to avoid any confusion with the CDW amplitude and phase channels).
Also included are the fitted spectra from the mode analysis (based on modes at the frequencies $\nuzn(T)$ denoted by the vertical lines).
One can resolve modes extending out to $9\THz$, which can be fitted both in terms of their spectral amplitude and phase, with a clear broadening of the features with increasing $T$. One sees how numerous modes manifest in $|S(\nu)|$ not as symmetric peaks, but rather with a derivative-like structure or destructive dips in the cumulative background of the other modes.
Such features clearly hamper approaches to fit the spectra on the basis of $|S(\nu)|$ alone, but are handled naturally by the inclusion of the spectral phase in the analysis. Moreover, the correspondence between the experimental and fitted phase spectra in Fig.~\ref{fig:opop}(b) also provides additional support for the validity of the fitted mode spectra.

To examine these spectral structures more closely, in Fig.~\ref{fig:opop}(c) we show a magnified range for the modes in the range 2-3~THz for $T=10\Kel$.  One sees that we resolve a doublet of two narrow adjacent modes at 2.23 and 2.24~THz (as well as two relatively close modes at 2.56/2.61~THz).  The former doublet was previously analyzed by fitting a \textit{single} mode lineshape \cite{schaefer10,schaefer14} to $|S(\nu)|$ in 
that spectral region.
One sees that for both doublets, $|S(\nu)|$ shows a significant dip between the two adjacent modes.  In order to assess this, we also calculate the ``incoherent'' band sum spectrum, $\usr{S}{Ram}(\nu)$, i.e. corresponding to the intensity sum of each fitted mode lineshape $S_n(\nu)$, i.e. $\usr{S}{Ram}(\nu)=[\Sigma_n|S_n(\nu)|^2]^{1/2}$.  This is also included in Fig.~\ref{fig:opop}(c) (black curve), and corresponds to the signal one would measure in conventional spontaneous Raman scattering measurements (notwithstanding possibly different relative band strengths, due to the distinct matrix elements for spontaneous vs. impulsive Raman interaction \cite{stevens2002}).  Evidently, the incoherent spectrum does not exhibit such pronounced local minima between the bands (nor the derivative-like structures for the modes at 2.69 and 2.77~THz), further emphasizing how band interference arises and provides more detailed information for the coherent Raman approach here.
A cursory consideration of the spectral interference might lead one to conclude that the neighboring mode pairs are significantly out of phase.  
To address this, in Fig.~\ref{fig:opop}(d), we plot the real/imaginary parts ($S^{(n)}_{r,i}$) of the fitted mode-resolved spectra $S_n(\nu)$ (Eq.~\eqref{eq:opopband}) for the doublet modes.  An inspection of the real parts demonstrates that all modes have indeed a phase $\varphi_{0n}$ close to zero (corresponding to a displacive, cosine time dependence in coherent Raman pump-probe studies \cite{stevens2002}), i.e. each $S^{(n)}_{r}(\nu)$ 
corresponds to a peak with nearly symmetric shape,
while each $S^{(n)}_{i}$ possesses a derivative shape, as well known for Lorentzian profiles in spectral response functions.  Note that for $\varphi_{0n}\rightarrow \pm \pi/2$, the real and imaginary lineshapes would indeed exchange shapes, as known for the more general Fano lineshape \cite{fano61}.
An inspection of $S^{(n)}_{i}$ then clarifies why destructive interference is observed between the two bands, as one sees that these are inherently of opposite sign (while constructive interference indeed occurs at their respective peak frequencies $\nuzn$).
Correctly accounting for this effect is clearly vital, e.g. if one were to assess the relative phase of neighboring modes in terms of the intermediate spectral structure (e.g. two neighboring modes in \textit{anti}-phase would exhibit \textit{constructive} interference between the peaks).

While we defer a presentation and analysis of the mode frequencies and damping vs. $T$ to Sec.~\ref{sec:modes} (Fig.~\ref{fig:modes}), clearly we now have a much more comprehensive set of Raman-active modes (compared to the previous analyses, which concentrated on modes at 1.68, 2.22 and 2.55~THz), to assess as candidates for CDW AMs.
Nevertheless, as discussed in the Introduction, the presence of complementary PMs is decisive for an unequivocal assignment of these bands to collective CDW modes, as one could always consider that these are usual Raman-active phonon modes which arise purely from zone-folding in the CDW phase \cite{sagar_raman08}.  
Due to the centrosymmetry in \kmo, one expects Raman-IR exclusion in the selection rules 
for conventional phonons, such that the appearance of corresponding modes is compelling evidence for their assignment as CDW modes.

\section{Phase modes: Reflective THz time-domain spectroscopy}\label{sec:thz}

\begin{figure}[h!]
\centering
\includegraphics[width=\linewidth]{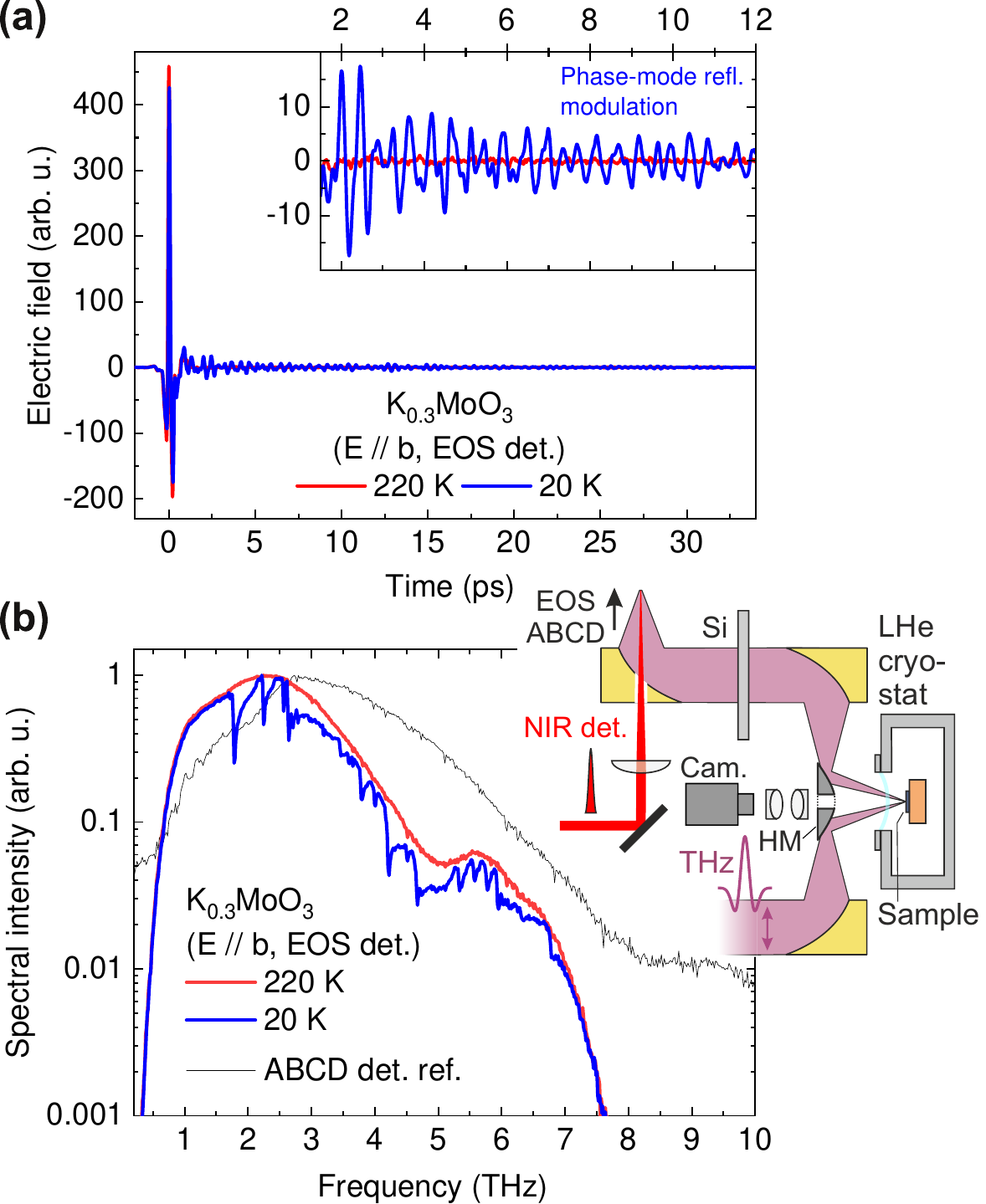}
\caption{(a) Example of detected THz time-domain fields with reflection sample geometry (and EOS detection): \kmo sample in the metallic phase $T=220\Kel>T_c$ (red curve, used as reference) and in the CDW phase $T=20\Kel \ll T_c$ (blue curve).  Inset shows magnified range of oscillatory signatures after the main pulse for the \kmo sample at low $T$ (while the weak residual oscillations for $T=220\Kel$ are due to residual water-vapor absorption in the THz beam path).
(b) Corresponding intensity spectra, including an additional reference spectrum using ABCD detection (see Sec.~\ref{sec:details}, used to provide extended bandwidth coverage for $T=20\Kel$). 
Also included is a schematic of a selected portion of the THz-TDS setup.
}
\label{fig:tds}
\end{figure}

\begin{figure*}
\centering
\includegraphics[width=\textwidth]{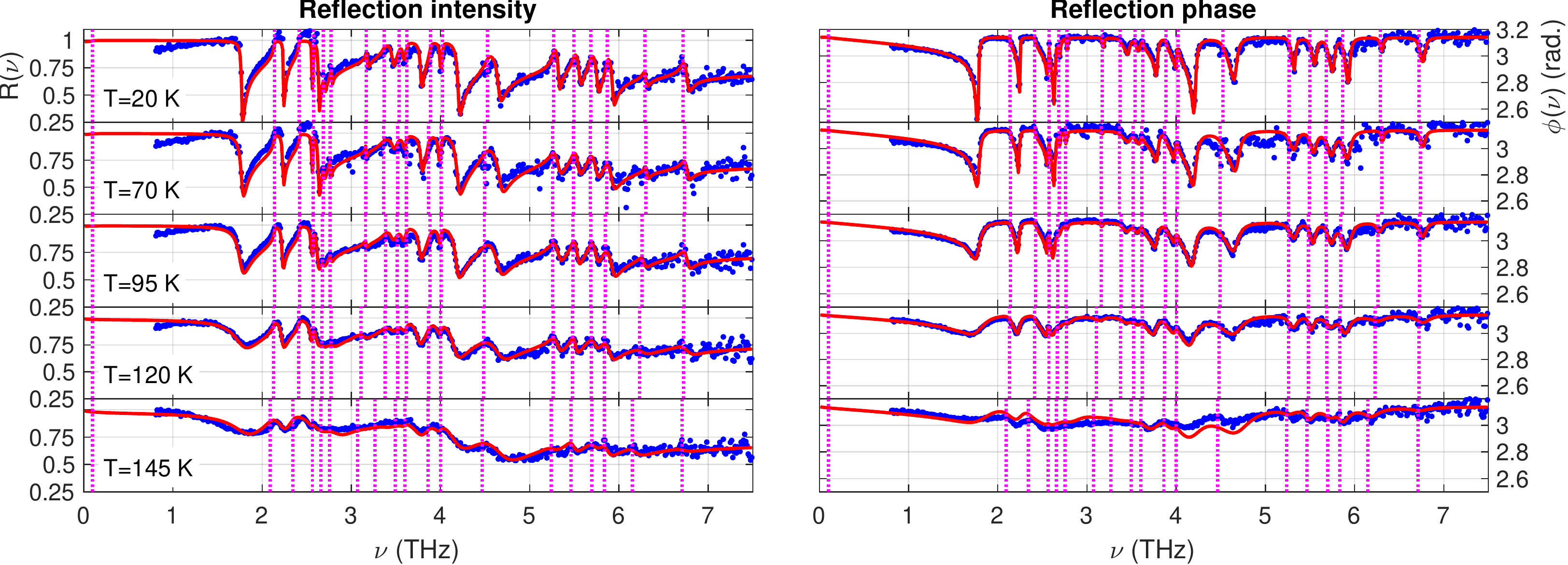}
\caption{Reflection spectral intensity ($R(\nu)$, left) and phase ($\varphi(\nu)$, right) obtained from THz-TDS measurements at the indicated temperatures (blue points), and fitted spectra (red curves).  Magenta vertical lines denote position of fitted modes $\nuzn$
(see text and Appendix~\ref{app:tds} for details on baseline correction method).  Lowest mode was fixed at the nominal value 0.1~THz for all $T$ during fitting. 
} \label{fig:tds_spec}
\end{figure*}

We now proceed to the PM results using reflection THz-TDS (see Sec.~\ref{sec:details} for details).
Examples of the detected THz pulse's temporal electric fields are shown in Fig.~\ref{fig:tds}(a), measured after reflection from the \kmo sample at both $T=220\Kel$ (in the metallic phase) and $T=20\Kel$ (in the CDW phase), with the 
corresponding intensity spectra in Fig.~\ref{fig:tds}(b) obtained by Fourier transformation.
One sees clearly the appearance of reflective dips across the spectrum for \kmo in the CDW phase.
Equivalently, this manifests in the time-domain field as a long oscillatory tail in the reflected field (while the main pulse is only weakly affected).
As discussed in Sec.~\ref{sec:details}, for these experiments we ensured that all additional reflections in the THz beam path are significantly delayed (or 
their effect minimized, as in the case of the thin polymer cryostat window, 
which produces weak internal reflections with a small temporal delay).
This allows a long time range and hence fine spectral resolution ($\simx 25$~GHz), without introducing spectral modulation from signal echos.
Despite efforts to obtain reference spectra with a gold mirror at the sample position, due to issues with the baseline (depending sensitively on alignment), 
we rather employ the \kmo sample in the metallic phase at $T=220\Kel>T_c$ as the reference, where one has a broad metallic response (for fields polarized along the b-axis, with $R_0$ varying smoothly in the range $0.8-0.9$) and negligible additional spectroscopic features in our measured THz frequency range \cite{beyer2012}.
The reflectivity spectra for a set of temperatures are shown in Fig.~\ref{fig:tds_spec}, both in terms of the (a) intensity $R(\nu)$ and (b) phase $\varphi(\nu)$. 
In contrast to the last section, where one obtains the mode spectra directly from the impulsive Raman signals, for the reflectivity measurements one must retrieve these via the complex Fresnel field coefficient -- see Appendix~\ref{app:tds} for details, including the baseline correction method used to account for the non-ideal reference.
The fitted reflection intensity- and phase-spectra are included in Fig.~\ref{fig:tds_spec}, based on a set of Lorentzian conductivity bands (Eq.~\eqref{eq:lorentz}), with mode frequencies $\nuzn$ denoted by vertical dotted lines (see Supplementary for fitted conductivity spectra, and next section for mode parameters vs. $T$).
One sees that the model reproduces both the reflection intensity- and phase-spectra well across the full bandwidth and, as per the last section, the inclusion of the spectral phase is decisive here to achieve a robust fit with numerous modes, especially with the mode broadening for increasing $T$. 

Due to the low-frequency roll-off in the spectral intensity (Fig.~\ref{fig:tds}(b)), we cannot perform a quantitative analysis of any modes in the region below 1~THz.  
Nevertheless, as shown in the Supplementary, a series of fitting tests with a low-frequency mode fixed at positions $\nu_{01}$ in the range 0-1.7~THz indicate that such a mode is indeed required, in order to obtain the reflectivity dip at $1.75\THz$ (most pronounced at low temperatures). This dip essentially manifests due to the interference of the tails of this low-frequency mode and the next higher one at 2.14~THz.
As mentioned in the Introduction, this feature previously led us to fit a PM very close to 1.75~THz, on the basis of non-equilibrium differential reflectivity spectra \cite{thomson17}.
The analysis of our ground-state spectra here results in a smaller misfit as  $\nu_{01}$ is lowered towards 1~THz, 
with the misfit then remaining essentially independent of $\nu_{01}$ for values below 1~THz.
Hence, we tentatively fixed the position of this low-frequency mode to $\nu_{01}=0.1$~THz, as per the experimentally proposed position of the ``phason'' in previous studies \cite{mihaly1989,degiorgi91}, for fitting our spectra for all $T$.
In order to provide additional data for PMs at higher frequencies than covered with EOS detection (data in Fig.~\ref{fig:tds_spec}), we also carried out an additional measurement at $T=20\Kel$ with extended bandwidth using ABCD detection (see Sec.~\ref{sec:details}, Fig.~\ref{fig:tds}(b), and Supplementary).

\section{Combined mode analysis: Time-dependent Ginzburg-Landau model}\label{sec:modes}

\begin{figure}[!h]
\centering
\includegraphics[width=0.85\linewidth]{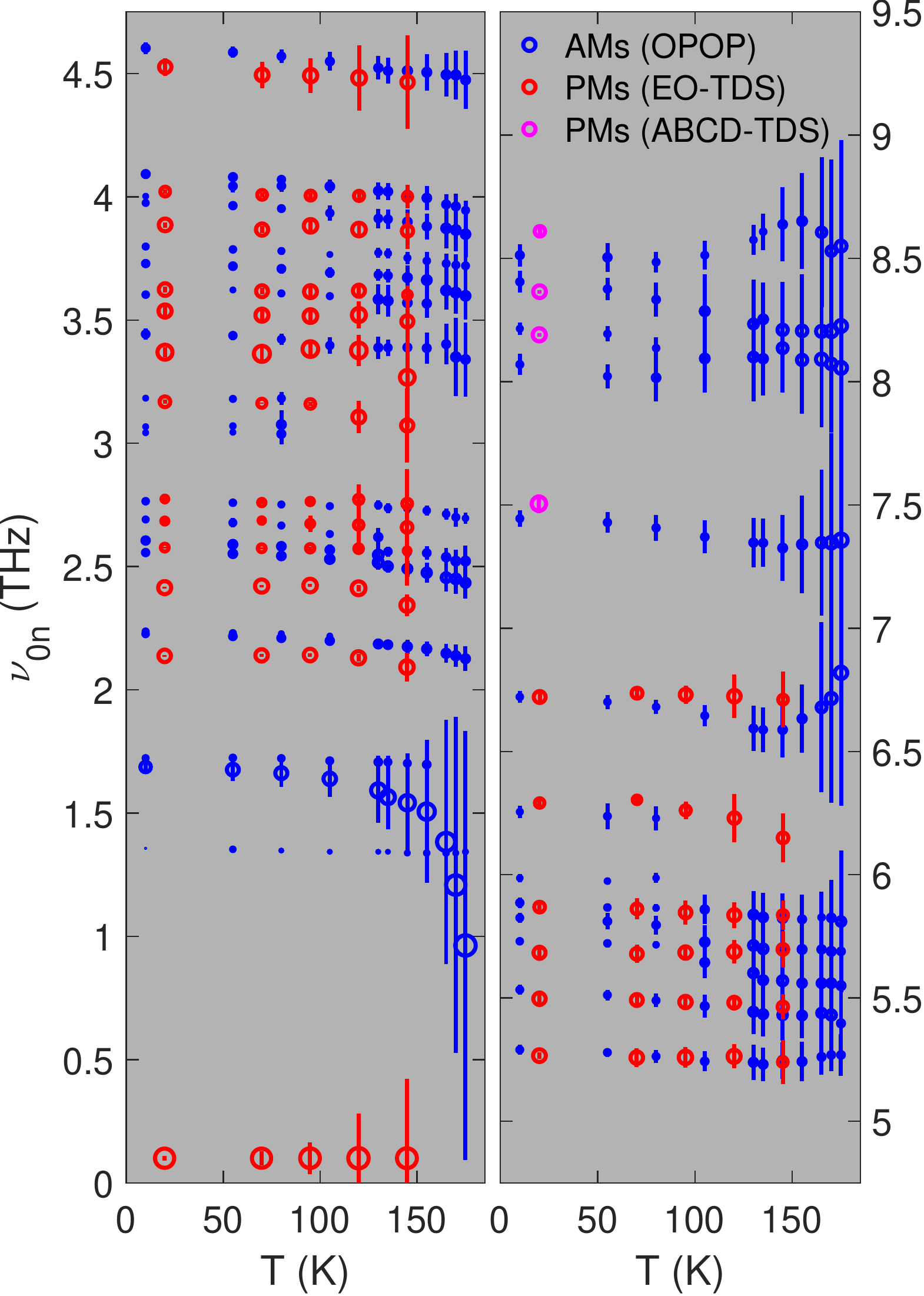}
\caption{
Combined temperature dependence of the experimental AMs (blue) and PMs (red) (left and right panels cover the lower and higher frequency ranges, respectively).  Vertical bars denote the band half-maximum widths $\pm\Gamma_n/2\pi$ centered on each mode frequency point \nuzn.  Additional PM data at $T=20\Kel$ (magenta) from extended-bandwidth (ABCD) THz detection (see Sec.~\ref{sec:details}).
} \label{fig:modes}
\end{figure}

In this section, we present the combined AM and PM results, and apply the TDGL model to substantiate their assignment as CDW collective modes and account for their $T$-dependence.
In Fig.~\ref{fig:modes}, we plot the fitted Raman-/IR-active mode frequencies $\nuzn(T)$ (from the last two sections, respectively), with the respective band-half-widths $\Gamma_n/2\pi$ denoted by vertical bars on each data point to depict the broadening.
One indeed observes a close pair-wise correspondence between the frequencies 
of the AMs and PMs (\nuznA and \nuznP, respectively), with the understanding that 
for the lowest-frequency respective modes, conventionally referred to as the ``amplitudon'' and ``phason'' \cite{gruener_book}, one expects $\nuzP\rarr 0$ (fixed in our spectral analysis at $\nuzP=0.1\THz$ \cite{mihaly1989,degiorgi91}, see Sec.~\ref{sec:thz}) 
and hence $\nuzA \gg \nuzP$ (with $\nuzA=1.68\THz$ at low $T$).
Significant broadening for $T\rarr T_c$ is observed for many of the modes, particularly so for the lowest AM (as reported before \cite{schaefer10,schaefer14}), but also significantly for the newly analyzed AMs above 6~THz.  While the available PM data does not approach $T_c$ as closely, the onset of a similar degree of broadening is also observed for most of the PMs for $T\rarr 145\Kel$. 
Compared to the previously reported AM analysis results \cite{schaefer10}, our new analysis approach here allows estimates of \nuzA approaching closer to $T_c$, and shows a more 
pronounced softening, with \nuzA falling to $\simx 1\THz$ at $T=175\Kel$.
Also, here we took care to fit the \textit{damped} mode frequencies $\wzn=2\pi\nuzn$ (see Eq.s~\eqref{eq:opopt} and \eqref{eq:opopband}), which are also those directly yielded from the TDGL eigenvalues below.  
As shown in the Supplementary, the new $\nuzA(T)$ data are also more consistent with those from conventional Raman \cite{travaglini_raman83,sagar_raman08} and neutron-diffraction \cite{pouget91} studies for $T\rarr T_c$. 

As all fitted modes exhibit a correspondence compatible with CDW collective modes, we apply a revised TDGL model with a bare coupled phonon (i.e. originally at $q=2\kF$ for $T>T_c$ with frequency $\Wzn$) for each experimental AM/PM pair (excluding the sharp, weak Raman-active modes at 1.36~THz, and at 1.72~THz just above the amplitudon \cite{schaefer10}).
The implementation of the TDGL is similar to that in our previous reports \cite{schaefer10,schaefer14,thomson17} (summarized again in the Supplementary), with the major difference here that we use a significantly weaker damping parameter $\gamma_2=0.09\cdot\gamma_1$ for the EOP-phase ($\Delta_2$) compared to $\gamma_1$ for the EOP-amplitude ($\Delta_1$), and hence also maintain the general-damping form (and not the overdamped limit \cite{schaefer10}) of the TDGL equations of motion for the phase channel. 
Note that the choice of equal nominal damping used in our previous PM study \cite{thomson17} followed certain assertions in the literature \cite{hennion92,tutis91}, the notion being that classically, the 
charge-density-compression/rarefaction (EOP-amplitude) and translation (EOP-phase) involves motion of the same condensate carriers. 
As mentioned above, the use of an equally strong damping for the EOP-phase (in combination with a strong phase-pinning parameter $\Wpin$) results in a lowest PM $\nuzP$ at higher frequencies, closer to $\nuzA$. 
However, such a prediction is no longer consistent with our revised determination of the lowest experimental PM.
There are indeed assertions in the literature that phase damping/relaxation can be significantly slower than for the amplitude in the case of CDW, where the quasi-particle excitations are neutral 
(contrary to the case in superconductivity with charged quasi-particles, where the magnitudes of the amplitude/phase damping rates are reversed) \cite{yusupov10,dolgirev20}. %
We note though, that these assertions are generally made in the context of the \textit{resulting} collective modes, while we instead consider here the appropriate, inherent damping magnitudes to be used for the EOP as input to the TDGL model, which in turn predicts the damping of the collective modes.
Nevertheless, as shown in the following, we find that the choice $\gamma_2=0.09\cdot\gamma_1$ does lead to revised TDGL predictions where the lowest PM is now close to zero-frequency (consistent with a nearly gapless phason), while providing a reasonable description of the other PMs.

\begin{figure*}
\centering
\includegraphics[width=\textwidth]{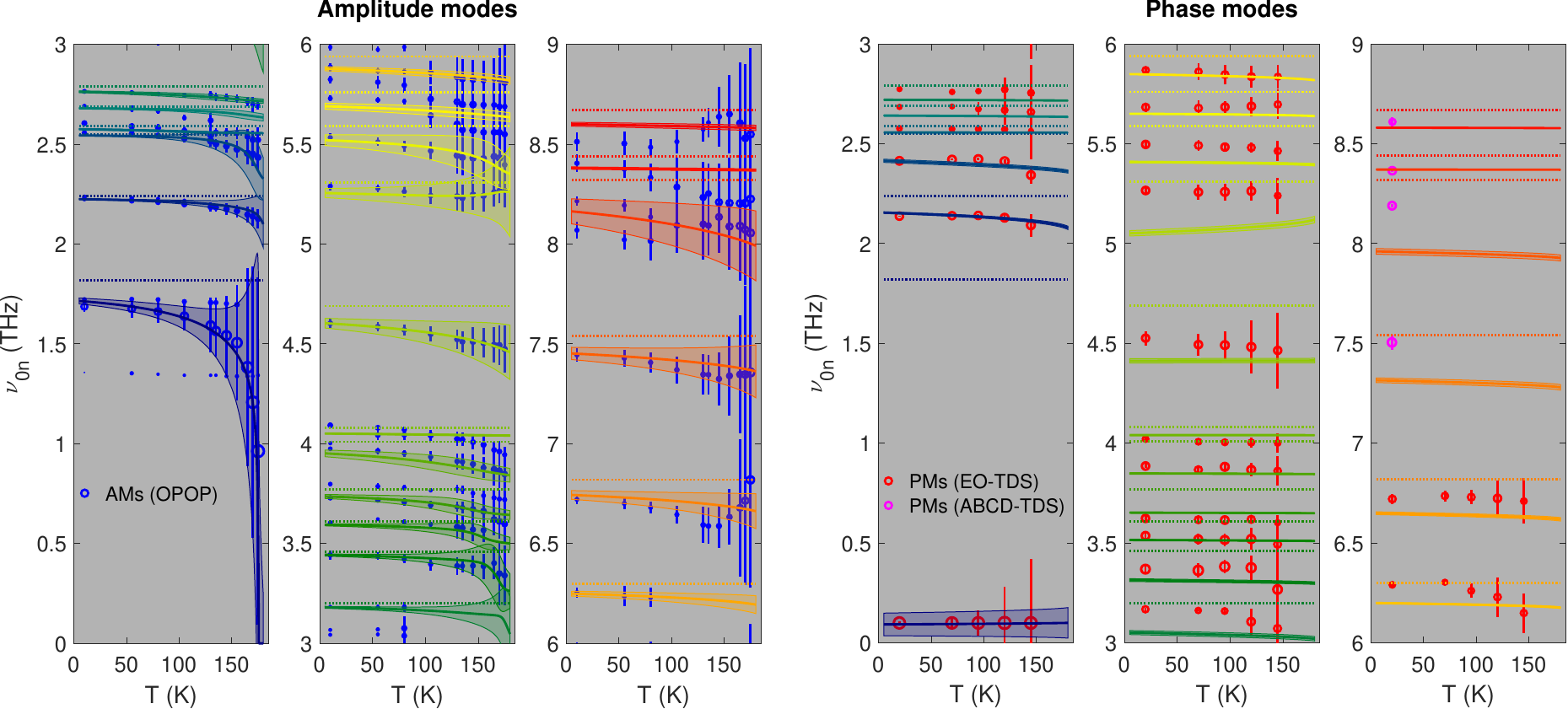}
\caption{
Comparison of AMs (left) and PMs (right) (Fig.~\ref{fig:modes})  with fitted TDGL model predictions. Equally colored fit curves correspond to one respective bare mode (shown as dashed lines).
} \label{fig:tdgl}
\end{figure*}

The results of the revised TDGL model are shown in Fig.~\ref{fig:tdgl}, where we plot the AMs and PMs separately for clarity, along with the experimental data from Fig.~\ref{fig:modes} (with each channel plotted in three graph columns to allow better inspection of each frequency range).
The predicted modes are plotted as filled regions tracing out $\nuzn(T)\pm\Gamma_n(T)/2\pi$ for direct comparison with the experimental mode frequencies/damping, while 
the bare phonon frequencies \Wzn are included as horizontal dashed lines (the mode coupling parameters $m_n$ and their bare-mode frequency dependence are discussed in detail below and presented in Fig.~\ref{fig:coupling}, while a full account of the other parameters is given in the Supplementary).  
To achieve these TDGL results, we tuned the bare mode frequencies $\Wzn$ and coupling parameters $m_n$, in conjunction with the global Ginzburg-Landau parameter $\alpha$ (which determines the restoring force about the $T$-dependent equilibrium EOP amplitude $\Delta_0=\alpha(T_c-T)/\beta$) and impurity pinning potential via $\Wpin$, in order to best reproduce both the experimental AMs and PMs simultaneously. 
(Note that $\beta$ is the coefficient of the quartic term $\propto \Delta^4$ in the TDGL potential energy, and drops out of the equations for the AM/PM frequencies at the equilibrium).
Here we employ a $T$-independent model for the impurity pinning $\Wpin$ (see Supplementary), which yields a pinned phason with nearly constant frequency, as observed experimentally \cite{mihaly1989,degiorgi91}.  
The model results show overall a good qualitative agreement with experiment, near-quantitatively for many of the modes, although certain systematic deviations are evident, such as the precise PM frequencies and some of the trends approaching $T_c$, in particular the lack of PM broadening which is significant for several experimental PMs (including the phason at 0.1~THz, although here the fitting of the bandwidth is tentative, given that the mode peak lies outside the fitting range). Still, such deviations are not surprising, considering the simplicity of the phenomenological TDGL model.

The assumption of $T$-independent coupling parameters $m_n$ neglects any influence of, e.g., the presence of normal electrons (density \usr{N}{th}) thermally promoted across the CDW gap as $T$ approaches $T_c$, which could screen the e-ph coupling \cite{rice1979,mihaly1989}. 
As the relative fraction of charges in the CDW condensate should follow 
$\usr{N}{C}(T)/N_0=\delta_0(T)$ 
(where $\delta_0(T)=\Delta_0(T)/\Delta_0(0)=\sqrt{1-T/T_c}$), in a two-fluid model we then have 
$\usr{N}{th}(T)/N_0=1-\delta_0(T)$, and allow for a $T$-dependent coupling via 
$m_n(T)=m_n(0)(1-b_n\cdot \usr{N}{th}(T)/N_0)$.
Indeed this correction (applied sparingly to selected bare phonons $\Wzn$) allowed us to refine the correspondence for the PMs between 2-2.5~THz (applied to the bare phonon at $1.82\THz$), and the   
AMs between 5-6~THz (applied to the bare phonons at 5.31, 5.59 and 5.76~THz), each with 
moderate values $b_n=0.3(5)$ (see Supplementary) -- these extensions are incorporated in the TDGL results in Fig.~\ref{fig:tdgl}.

We assessed incorporating several other, physically plausible $T$-dependent effects into the TDGL model, to see if these might readily account for the remaining deviations, as discussed in the following. %
To investigate mechanisms which could lead to PM broadening, we considered the effects of 
(i) $T$-dependent EOP-phase damping ($\gamma_2(T)$ increasing for $T\rarr T_c$) and (ii) inherent bare-phonon damping with thermal anharmonic broadening \cite{schaefer10}.  Neither of these extensions provided a convincing improvement for describing the experimental trends, where we observed that the bare-phonon damping does not translate directly to the resultant PM damping.
While such directions to extend the TDGL description deserve further investigation in ongoing studies, it seems prudent to first develop an estimate of the expected magnitude of such corrections from microscopic models, before incorporating them in the phenomenological TDGL framework here.

To conclude this section, we focus on the magnitude of the coupling parameters $m_n$, in particular their dependence on their respective bare-mode frequencies $\Wzn$.
Within the TDGL model, at equilibrium, the amplitude of the $n$th phonon coordinate is given by
$\xi_{0n}=(m_n/\Wzn^2)\Delta_0$, 
which results in an elastic deformation energy cost of 
$U_{Ln}=+\tfrac{1}{2}\Wzn^2\xi_{0n}^2=\tfrac{1}{2}m_n^2\Delta_{0}^2/\Wzn^2$ but a stabilization energy of
$U_{Cn}=-m_n\Delta_0\cdot\xi_{0n}=-2 U_{Ln}$, i.e. twice the magnitude of the elastic energy cost.
Based on this $1/\Wzn^2$-dependence, one might infer that the contribution of each bare phonon to the CDW formation decreases with increasing $\Wzn$. However, we show here that, based on our TDGL parameters for \kmo, this effect is actually countered by the growth of $m_n$ vs. $\Wzn$ for the higher-energy phonons.

In Fig.~\ref{fig:coupling}, we plot $m_n$ vs. $\Wzn$ for the set of bare phonons employed in the TDGL analysis in Fig.~\ref{fig:tdgl}. 
As can be seen, there is a clear increasing trend vs. $\Wzn$.  To interpret this result more physically, one must transform the TDGL parameters to a measure which reflects the inherent e-ph coupling strength, as is the case for the dimensionless e-ph coupling parameters $\lambda_n$ in quantum-mechanical models \cite{rice76,rice78} (as was used tentatively in an early report of the phase modes in \kmo at a single temperature, $T=6\Kel$ \cite{degiorgi91}).
To this end, in Appendix~\ref{app:qm}, we derive a correspondence between $m_n$ and $\lambda_n$ (Eq.~\eqref{eq:lambda}), with the result that 
$\lambda_n\propto (m_n/\Wzn)^2$.  The relative calculated values of $\lambda_n$ are also shown in Fig.~\ref{fig:coupling}.  Ones sees that while the value of $\lambda_1$ indeed is significantly larger than the values of the subsequent modes, for $n\geq 2$ there is clear (roughly linear) increase in the dimensionless e-ph coupling, even after correcting for the inherent $\Wzn$-dependence in $m_n$.
This is in contrast to the treatment in \cite{degiorgi91}, where a constant nominal value of $\lambda_{n\geq 2}$ was assumed for the modes, and indicates that these stiffer phonons possess character which influence the electronic energy more significantly.  This result strongly motivates ab initio/DFT calculations to assign the structural character of the bare modes and investigate how they interact with the electronic orbitals in more detail.

\begin{figure}
\centering
\includegraphics[width=0.9\linewidth]{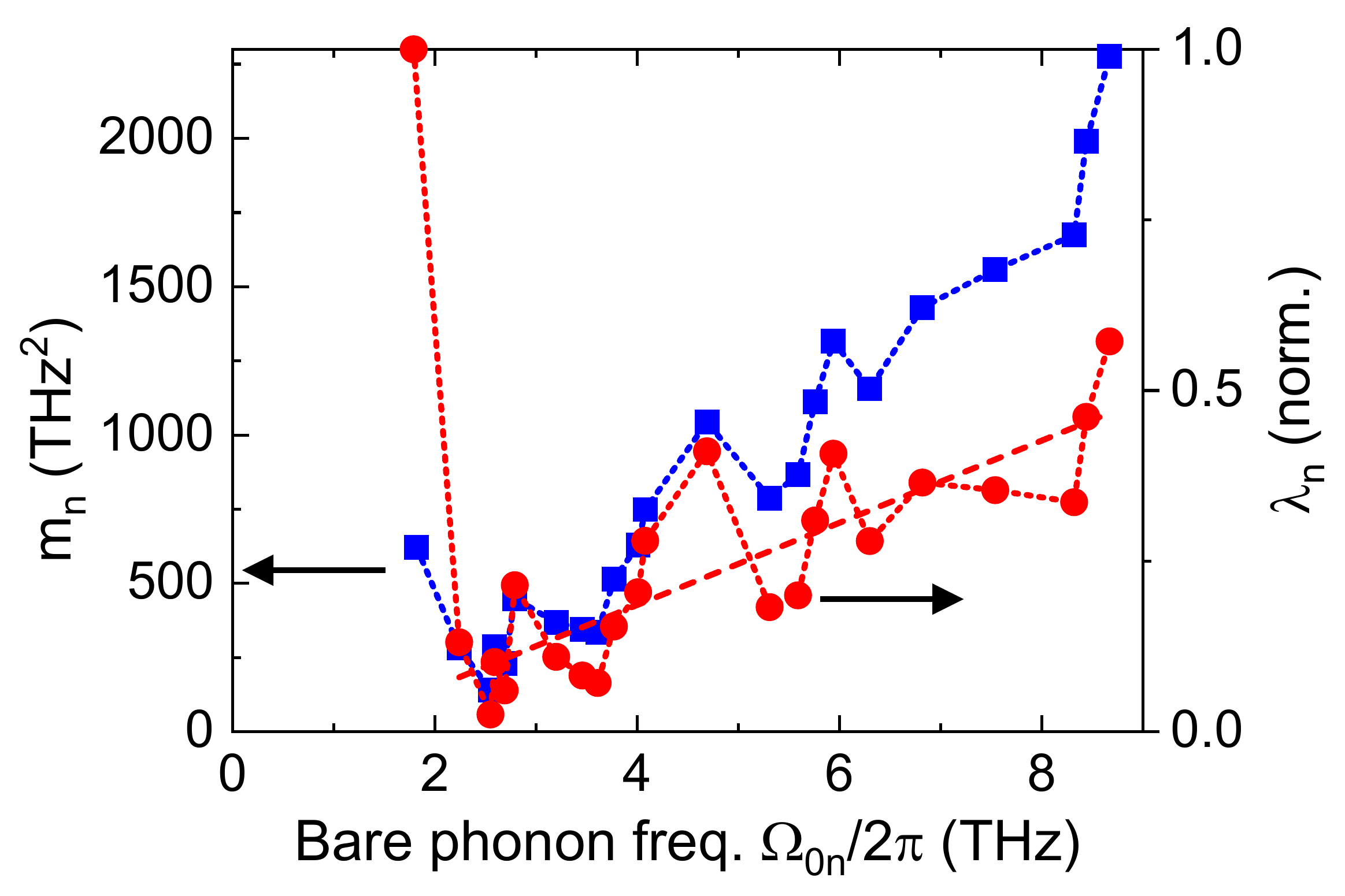}
\caption{
Dependence of e-ph coupling parameters from TDGL analysis on bare-mode frequency: (a) $m_n$ parameters from TDGL model (at $T=0$ for modes where $T$-dependence was employed), (b) relative dimensionless e-ph coupling parameters $\lambda_n$ accounting for inherent frequency scaling of $m_n$.
} \label{fig:coupling}
\end{figure}

\section{Conclusion}

The combined study of the CDW collective modes in \kmo for both amplitude and phase channels provides strong support for their assignment, whereby the TDGL model applied here indicates that higher frequency modes are indeed strongly coupled to the electronic density wave and play an important role in promoting the CDW phase.
These results strongly motivate first-principles calculations of the phonons and e-ph coupling, although this remains challenging for relatively complex materials such as \kmo.
From the experimental side, in addition to returning to the non-equilibrium response of these modes \cite{thomson17,rabia19}, a rigorous determination of the ground-state phase response in the low-frequency range (${\lesssim} 1.5\THz$) is still lacking, being complicated by the inherent issue of resolving spectral features in reflection with $R\approx 1$.  Here we are pursuing THz transmission studies of thin exfoliated flakes, although the small lateral dimensions of sufficiently thin samples (thickness ${<}10\mum$) are limited, requiring specialized spectroscopic methods.
Here deposited \kmo thin films \cite{staresinic2012,dominko2021}, which exhibit near-crystalline AM response, could provide essential experimental results, if their morphological properties maintain the PM response for macroscopic field interaction.
The TDGL model can be readily applied to account for the collective modes, and serves as a versatile framework which can be applied to systems with multiple, coupled order parameters and for ultrafast non-equilibrium studies \cite{yusupov10,hinton13,zhou21}. However, its phenomenological basis necessitates further studies based on microscopic quantum-mechanical/many-body models \cite{rice76,rice78} to better establish its validity and estimate effective parameters.
Current efforts here require the extension of such quantum-mechanical treatments to rigorously treat the finite-temperature case and account for effects due to e.g. Coulomb interactions and impurities, where developments have already begun \cite{hansen23}.

\section*{Acknowledgments}

We gratefully acknowledge funding by the German Research Foundation (DFG) via the Collaborative Research Center TRR 288 (422213477, project B08).  We thank Viktor Kabanov, Max Hansen, Yash Palan, Viktor Hahn, Falko Pientka, Oleksandr Tsyplyatyev, and Peter Kopietz for helpful discussions, and Virna Kisi\v{c}ek for participating in preliminary experiments.

\appendix
\section{Spectral analysis of impulsive Raman signals}
\label{app:opop}

The ansatz for the model differential reflectivity signal $S(t)\equiv \Delta R(t)/R_0$ vs. delay time $t$ following the pump pulse is given by:
\beq
S(t)=A_n e^{-t/\tau_n}\cos(\wzn t - \phin)\Theta(t)
\label{eq:opopt}
\eeq
(the sum over mode index $n$ is implied, $\Theta$ the Heaviside function), where incoherent (i.e. purely electronic) signal components correspond to $\wzn=0$.  Fourier transformation of Eq.~\eqref{eq:opopt} yields the sum over a set of general Lorentzian (Fano) lineshapes
\begin{flalign}
S(\omega) &= \frac
{(\Gamma_n + i\omega)A_n \cos\phin + \wzn A_n \sin\phin}
{(\wzn^2+\Gamn^2)-\omega^2+2i\Gamn\omega}  \notag \\
&=c_n f_n(\omega) + s_n g_n(\omega),
\end{flalign}
with damping $\Gamn=\tau_n^{-1}$, $c_n=A_n \cos\phin$, $d_n=A_n \sin\phin$, and the basis functions are given by 
\begin{flalign}
f_n(\omega)&=(\Gamn+i\omega)/D_n(\omega), \qquad    
g_n(\omega)=\wzn/D_n(\omega),  \notag \\
D_n(\omega)&=(\wzn^2+\Gamn^2)-\omega^2+2i\Gamn\omega.
\label{eq:opopband}
\end{flalign}
Clearly $A_n^2=c_n^2+s_n^2$ and $\tan\phin=s_n/c_n$.
For the purely electronic components ($\wzn=\phin=0$), this reduces to the Drude form $S_n=A_n/(\Gamn+i\wzn)$.

Taking into account the experimental impulse response $H(t)$ (i.e. 
cross-correlation of the pump- and probe-pulse intensity profiles, taken as a Gaussian function with FWHM temporal width of $\simx 80$~fs here), one has $S(t)\rightarrow S(t)\ast H(t)$, or for the spectra, $S(\omega)\rightarrow S(\omega)\cdot H(\omega)$, such that this response can be simply multiplied into the basis functions $f_n, g_n$.  For each iteration in the optimization algorithm, one generates the basis functions for the current values of $\wzn$ and $\Gamn$, and performs linear regression to minimize the misfit $\Sigma_j |S_j-\hat{S}_j|^2$, where $S_j=S(\omega_j)$ and $\hat{S}$ denotes the complex experimental spectrum. 

As discussed in the main text, while this approach allows one to simultaneously fit both the incoherent ($\wzn=0$)  and oscillatory components ($\usr{S}{el}$, $\usr{S}{osc}$, respectively), in practice for the spectra here we rather first perform a fit of $\usr{S}{el}$ in the time domain, and subtract this result to fit the modes in the residual spectrum
$\usr{S}{osc}=S-\usr{S}{el}$.  This allows one to closely scrutinize (and take steps to further minimize) any residual from the electronic response before fitting the modes, which is particularly important to cleanly fit the higher-frequency modes.

While fitting the complex spectra may appear to be equivalent to fitting the time-domain signal directly, the essential difference is that any deviations from the ideal model signal in Eq.~\eqref{eq:opopt} (e.g. due to frequency chirp or a non-exponential decay envelope) are more robustly ameliorated by the spectral misfit function.
% Add note on NL search algorithms?
% We typically used a robust Nelder-Mead simplex algorithm, although the ...

\section{Spectral analysis of THz reflectivity spectra}\label{app:tds}

For THz-TDS reflectivity measurements, one must retrieve the complex relative permittivity $\epsr{r}(\nu)$ from the complex reflectivity field coefficient $r(\nu)$, which, for our case of oblique incidence ($\theta=28^\circ$) and p-polarised field, is given by \cite{dresselbook}:
\beq
r=\sqrt{R}\cdot e^{i\phi}=
-\frac{\epsr{r} C_i-C_t}{\epsr{r} C_i+C_t},
\label{eq:fresnel}
\eeq
where $C_i=\cos{\theta}$ and $C_t=(\epsr{r}-\sin^2{\theta})^{1/2}$.
The corresponding conductivity spectrum is then calculated via 
$\sigma=i\omega\epsilon_0(\epsr{r}-\epsr{\infty r})$ \cite{hangyo05,thomson18}.
Due to the sensitivity of the directly recovered conductivity spectra to the precise reference baseline, especially here with strongly reflecting samples (as well as an inadvertent reference pulse delay $\delta t$ which introduces a phase term $r\rarr r\cdot e^{-i\omega\delta t}$) we instead fit $r(\omega)$ directly, with a conductivity model comprising a standard Lorentzian band for each mode \cite{dresselbook},
\beq
\sigma(\omega) = \frac{i\omega \sigma_{0n}}{\wzn^2-\omega^2+i\Gamma_n\omega}.
\label{eq:lorentz}
\eeq
(again summing over mode index $n$) with $\epsr{\infty r}$ and $\delta t$ included in the fit parameters to minimise the misfit $\Sigma_j|r_j-\hat{r}_j|^2$ to the experimental data $\hat{r}_j=\hat{r}(\omega_j)$.  
In order to compensate the resulting intensity/phase baseline of the non-ideal reference (the metallic phase of the \kmo sample), for each iteration we applied a complex correction factor $r(\omega) \rarr P(i\omega)\cdot r(\omega)$ where $P$ was taken as a third-order complex polynomial determined adaptively via regression of the model reflectivity spectrum.
The experimental results in Fig.~\ref{fig:tds_spec} correspond to those following this correction, i.e. $\hat{r}(\omega) \rarr \hat{r}(\omega)/P(i\omega)$.
While this approach introduces additional uncertainty into the fitting analysis, it still maintains a reasonable degree of robustness as one fits the full complex spectra (Fig.~\ref{fig:tds_spec}).
\\
\section{Correspondence of coupling parameters between TDGL and quantum-mechanical models}
\label{app:qm}

In Sec.~\ref{sec:modes} we obtain estimates of the coupling parameters $m_n$ for each coupled phonon mode (with bare frequencies \Wzn) from fitting the experimental modes with the TDGL model, as presented in Fig.~\ref{fig:coupling}.
In order to obtain parameters with a more transparent physical interpretation, here we derive a correspondence between the TDGL $m_n$ parameters and the dimensionless e-ph coupling parameters $\lambda_n$ from quantum-mechanical models \cite{rice76,rice78,khomskii_book}, albeit in a simplified classical limit,  
to account for any inherent dependence on $\Wzn$.
Following the development in \cite{khomskii_book}, one can write the lattice displacement along the $n$th phonon coordinates with wavevector $q=2k_F$ as 
$u_n=(2M_n\Wzn/\hbar)^{-1}\cdot 2b$ ($M_n$ the reduced mass), where 
one takes $(b_{nq}^{\dag}+b_{n,-q})\rarr 2b_n \delta(q-2 k_F)$ for the phonon operators.  
The elastic deformation energy cost is then $U_{Ln}=\tfrac{1}{2}M_n \Wzn^2u_n^2=\hbar \Wzn b_n^2$ while the coupling energy is 
$U_{Cn}=-2g_n \rho b_n$, where $g\equiv g_{2k_F}$ is the e-ph constant in the Fröhlich coupling term, and 
$\rho \equiv \rho_{2k_F}=\langle \Sigma_{k,\sigma}c^{\dag}_{k-2k_F,\sigma}c_{k,\sigma}\rangle$ represents the electronic density modulation amplitude.

The corresponding expressions based on the TDGL potential energy \cite{schaefer10,schaefer14,thomson17} are 
$U_{Ln}=\tfrac{1}{2}\Wzn^2 \xi_n^2$ and $U_{Cn}=-m_n\Delta \cdot \xi_n$.  Hence we can associate $\xi_n = \sqrt{M}\cdot u_n$ and
\beq
g_n=m_n \sqrt{\frac{\hbar}{2\Wzn}}\frac{\Delta}{\rho}.
\eeq
Due to the arbitrary scaling of the EOP amplitude $\Delta$ in the TDGL equations, the ratio $\Delta/\rho$ is undetermined but constant.
One then obtains for the dimensionless electron phonon constant:
\beq
\lambda_n = \frac{g_n^2N_0}{\hbar \Wzn} = \frac{N_0}{2}
\left(\frac{\Delta}{\rho}\right)^2\frac{m_n^2}{\Wzn^2},
\label{eq:lambda}
\eeq
where $N_0$ is the electronic density of states at the Fermi energy (in the normal undistorted phase).  Eq.~\eqref{eq:lambda} shows that the inherent e-ph coupling depends on the ratio $m_n^2/\Wzn^2$, which we then take into account in Sec.~\ref{sec:modes} (Fig.~\ref{fig:coupling}) in assessing the dependence on $\Wzn$.

%\bibliography{shorttitles,references,refs_extra}
%\bibliography{CPA_Lab,refs_extra}

\end{document}

\title{Supplementary Material to: ``Combined investigation of collective amplitude and phase modes in a quasi-one-dimensional charge-density-wave system over a wide spectral range''}

\author{Konstantin Warawa}
\affiliation{\ffm}
\author{Nicolas Christophel}
\affiliation{\ffm}
\author{Sergei Sobolev}
\affiliation{\jgu}
\author{Jure Demsar}
\affiliation{\jgu}
\author{Hartmut G. Roskos}
\affiliation{\ffm}
\author{Mark D. Thomson}
\affiliation{\ffm}
%\date{January 2023}

\maketitle

\newpage

\section{Amplitude modes: Impulsive-Raman all-optical pump-probe spectroscopy}
\label{sec:opopt}

\subsection{Time-domain signals and fitted incoherent signal components}

\begin{figure}[!h]
	\includegraphics[width=0.5\textwidth]{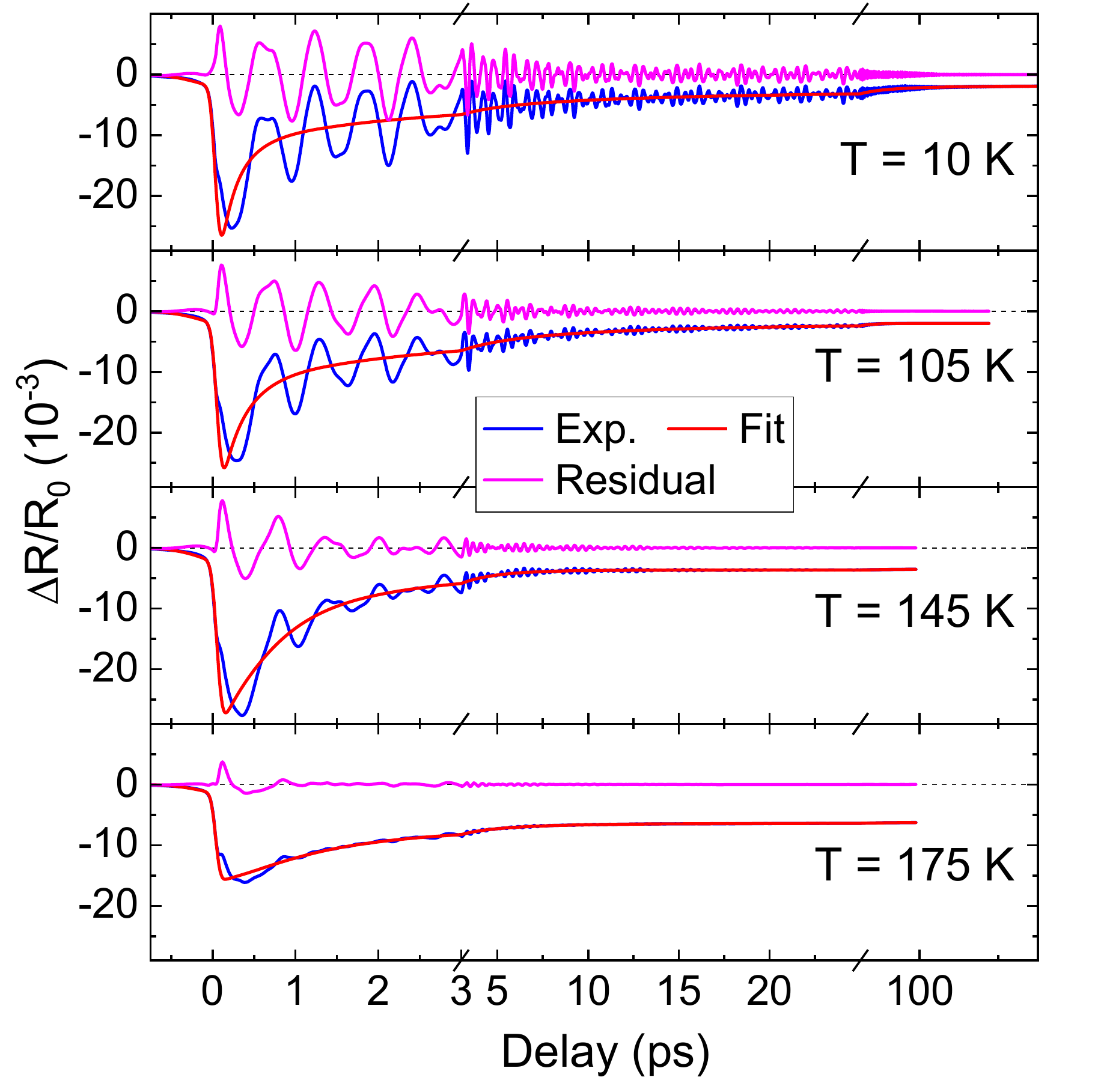}
\caption{
Time-domain differential reflectivity from all-optical pump-probe (impulsive Raman) experiments for selected temperatures, including fits to incoherent (predominantly electronic) signal components, and oscillatory residual, the latter used to analyze the amplitude-mode (AM) spectra in main paper.
} \label{fig:opopt}
\end{figure}
\begin{figure}[!h]
	\includegraphics[width=0.65\textwidth]{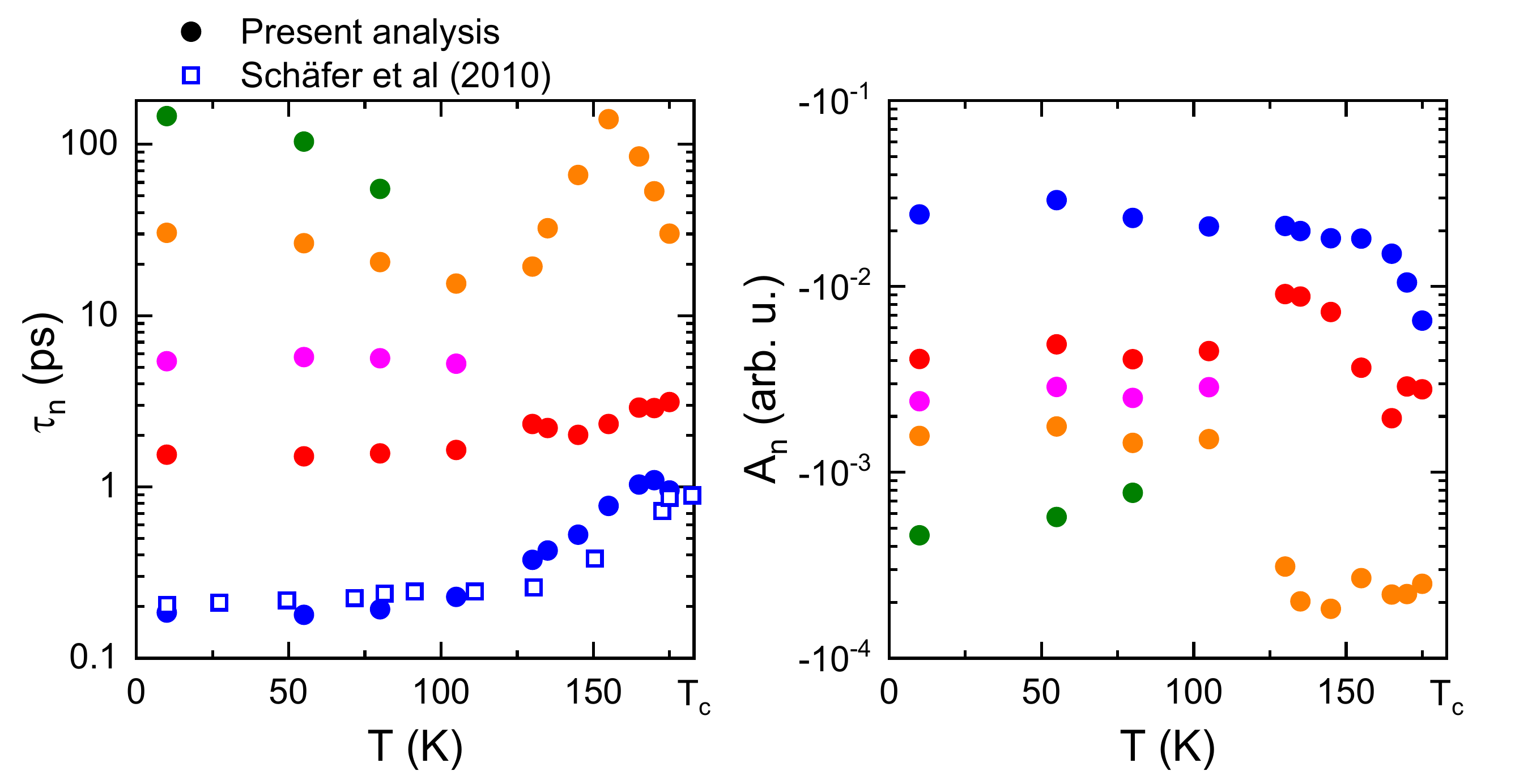}
\caption{
(a) Fitted exponential decay time constants and (b) corresponding amplitudes for incoherent (predominantly electronic) signal components vs. temperature (see Fig.~\ref{fig:opopt} for example kinetics).  Also included are the data for the shortest decay time $\tau_1$ from our previous report \cite{schaefer10}.  Common color coding used in both figures (e.g. the component with $\tau_1$ in (a) has the largest magnitude for the amplitude $|A_1|$ in (b)).
} \label{fig:opopt2}
\end{figure}

\subsection{Fitted lowest amplitude mode vs. literature reports}

\begin{figure}[!h]
	\includegraphics[width=0.7\textwidth]{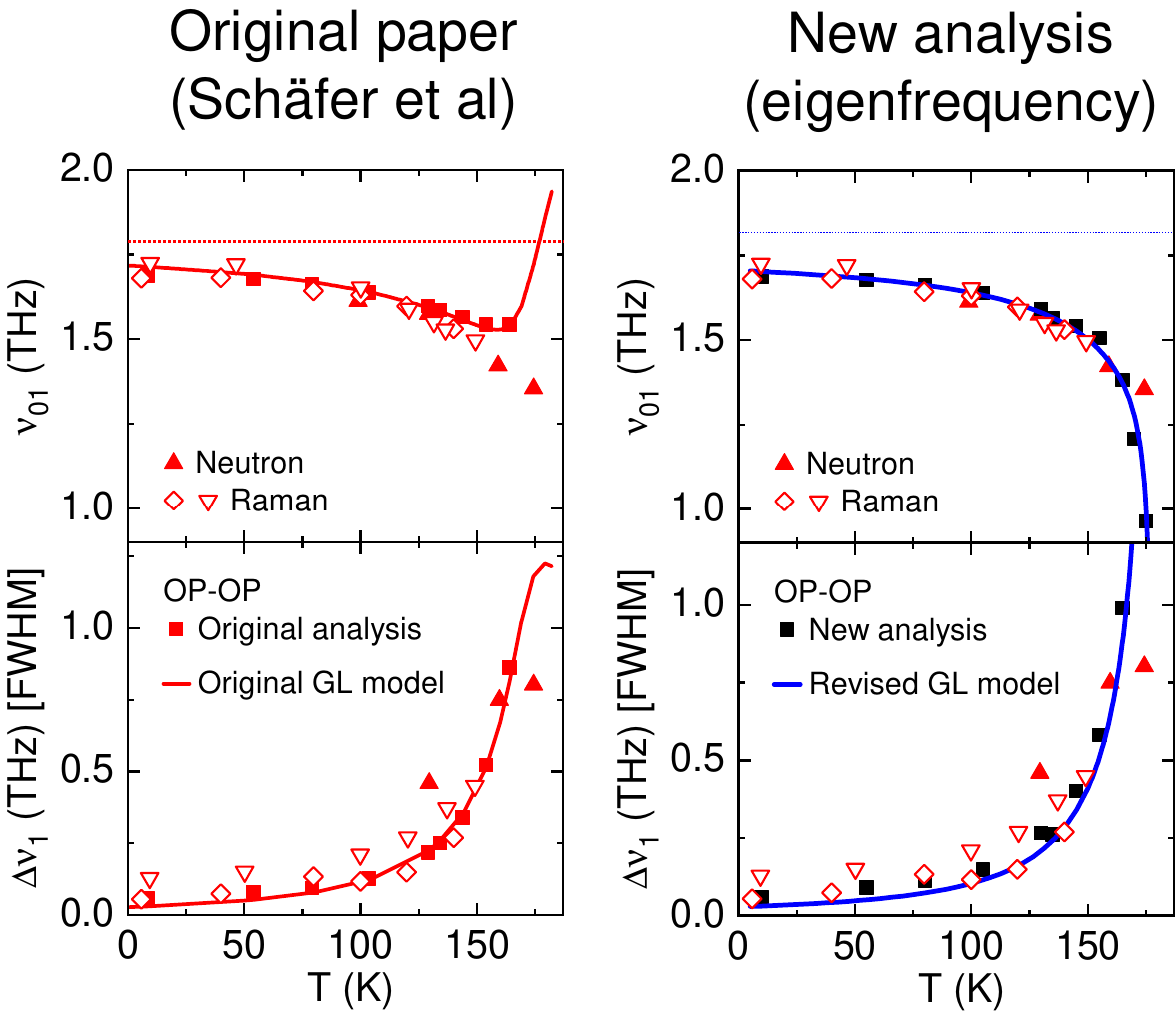}
\caption{Comparison of the fitted temperature dependence of the first AM frequency $\nuzA$ (``$\nu_{01}$'') and full-width half maximum (FWHM) bandwidth $\Delta\nu_1$ obtained from different experimental studies, including neutron diffraction ($\blacktriangledown$) \cite{pouget91}, Raman scattering ($\triangledown$, $\diamond$) \cite{travaglini_raman83, sagar_raman08} and all-optical pump-probe ($\blacksquare$) \cite{schaefer10}. The original analysis results from \cite{schaefer10} including a TDGL fit (left) are compared with those from the current paper (right), with the latter showing stronger mode softening and broadening approaching $T_c$ and better agreement with neutron and Raman data.
} \label{fig:schaefercomp}
\end{figure}

\newpage

\section{Phase modes: Reflective THz time-domain spectroscopy}\label{sec:thz}

\subsection{Fitted conductivity spectra}

\begin{figure}[htbp]
    \includegraphics[width=1\textwidth]{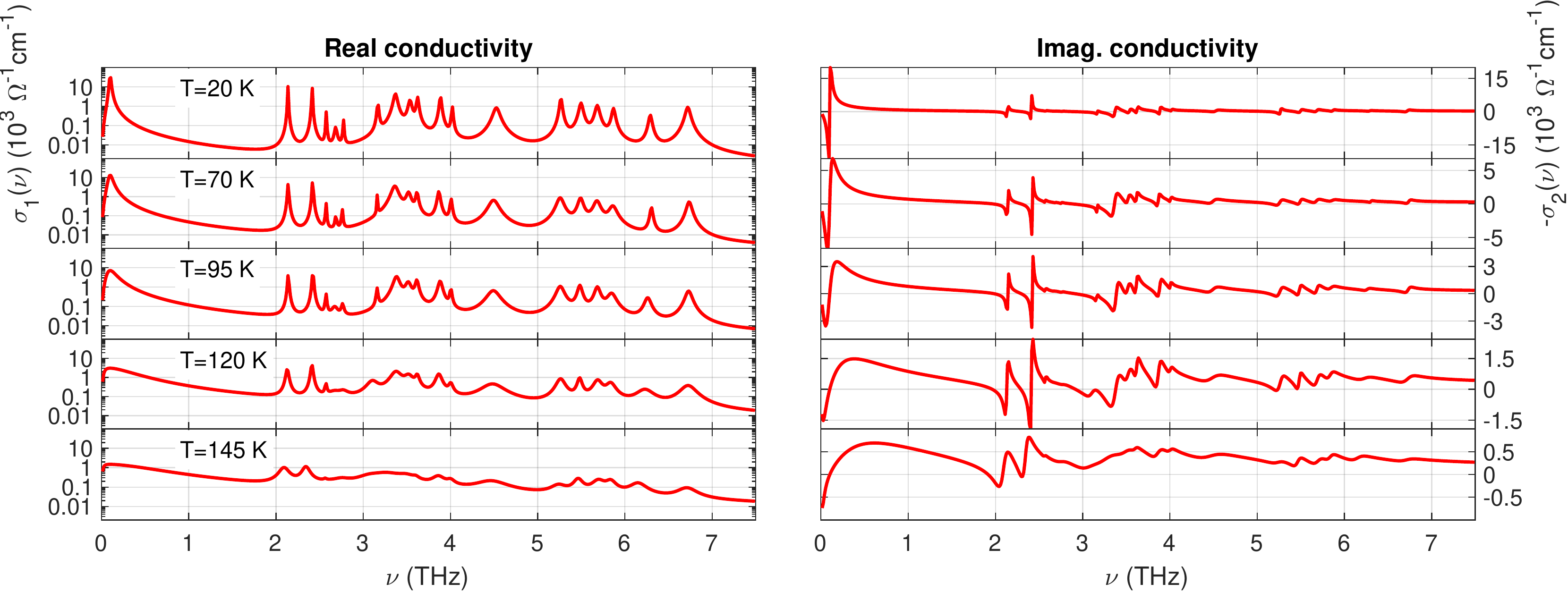}
\caption{Fitted complex conductivity spectra ($\sigma(\nu)=\sigma_1(\nu) + i\cdot\sigma_2(\nu)$) from THz reflectivity spectral analysis for each temperature $T$ (Fig. 3 in main paper). For clarity the real part $\sigma_1$ is plotted with a logarithmic vertical scale.
} \label{fig:fitsig}
\end{figure}

As described in detail in Sec.~IV and Appendix B of the main paper, we fitted the complex reflectivity spectra $r(\nu) = \sqrt{R(\nu)}\cdot e^{i\phi(\nu)}$ using a set of Lorentzian bands for the conductivity $\sigma(\nu)$
(which, due to the sensitivity to the precise reflectivity baseline, could not be inverted directly from $r(\nu)$ via the Fresnel reflection formula).
For completeness, we show these fitted conductivity spectra in Fig.~\ref{fig:fitsig}, where one can more readily visualize the relative band strengths and broadening with increasing $T$. 

\subsection{Fitting of lowest phase modes}

\begin{figure}[!t]
	\includegraphics[width=1\textwidth]{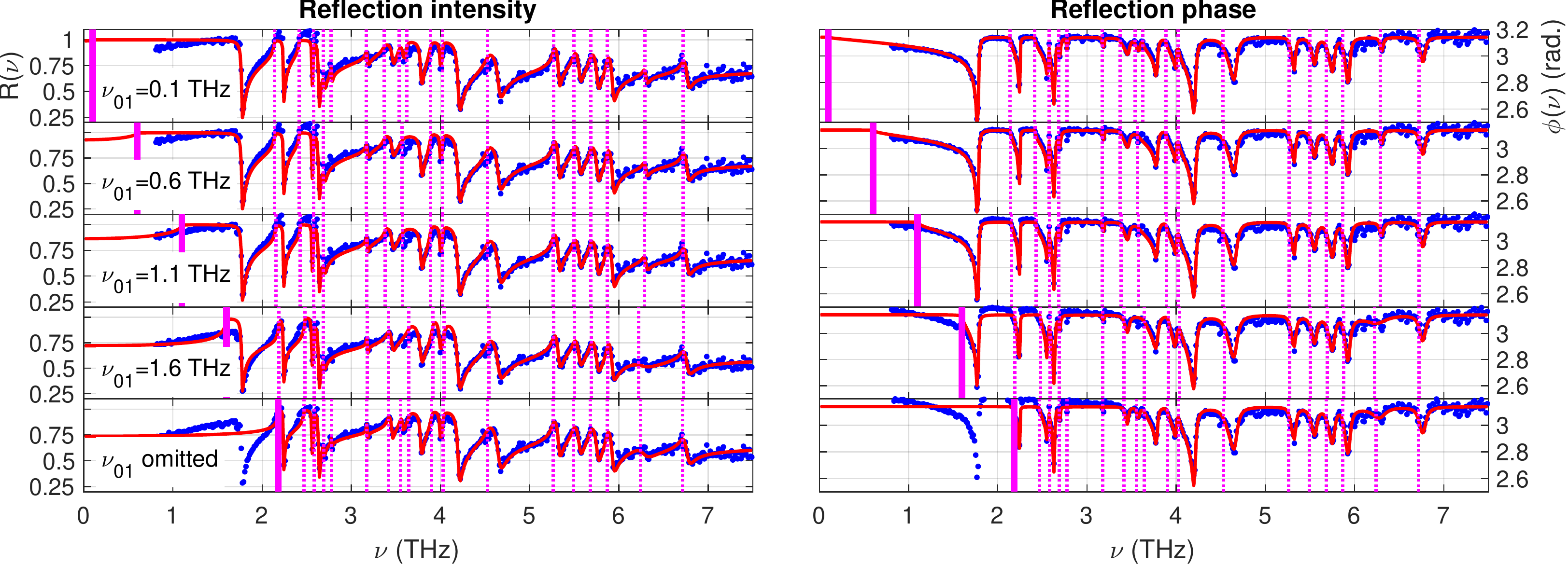}
\caption{Reflection spectral intensities ($R(\nu$)) and phases ($\varphi(\nu)$) obtained from THz-TDS measurements at $T=20\Kel$ (blue points), and fitted spectra for selected values of a fixed lowest-mode frequencies $\nu_{01}$ (red curves, as per the labels in the left column). Magenta vertical lines denote position of fitted modes $\nuzn$.
The first plot with $\nu_{01} = 0.1\THz$ is equivalent to the ``$T=20\Kel$''-plot in Fig. 3 from the main paper. A fit with no mode used below $1.75\THz$ is also shown (``$\nu_{01}$ omitted''), which cannot reproduce the strong reflectivity dip at that frequency.
} \label{fig:fixnu1spec}
\end{figure}

\begin{figure}[!t]
	\includegraphics[width=0.5\textwidth]{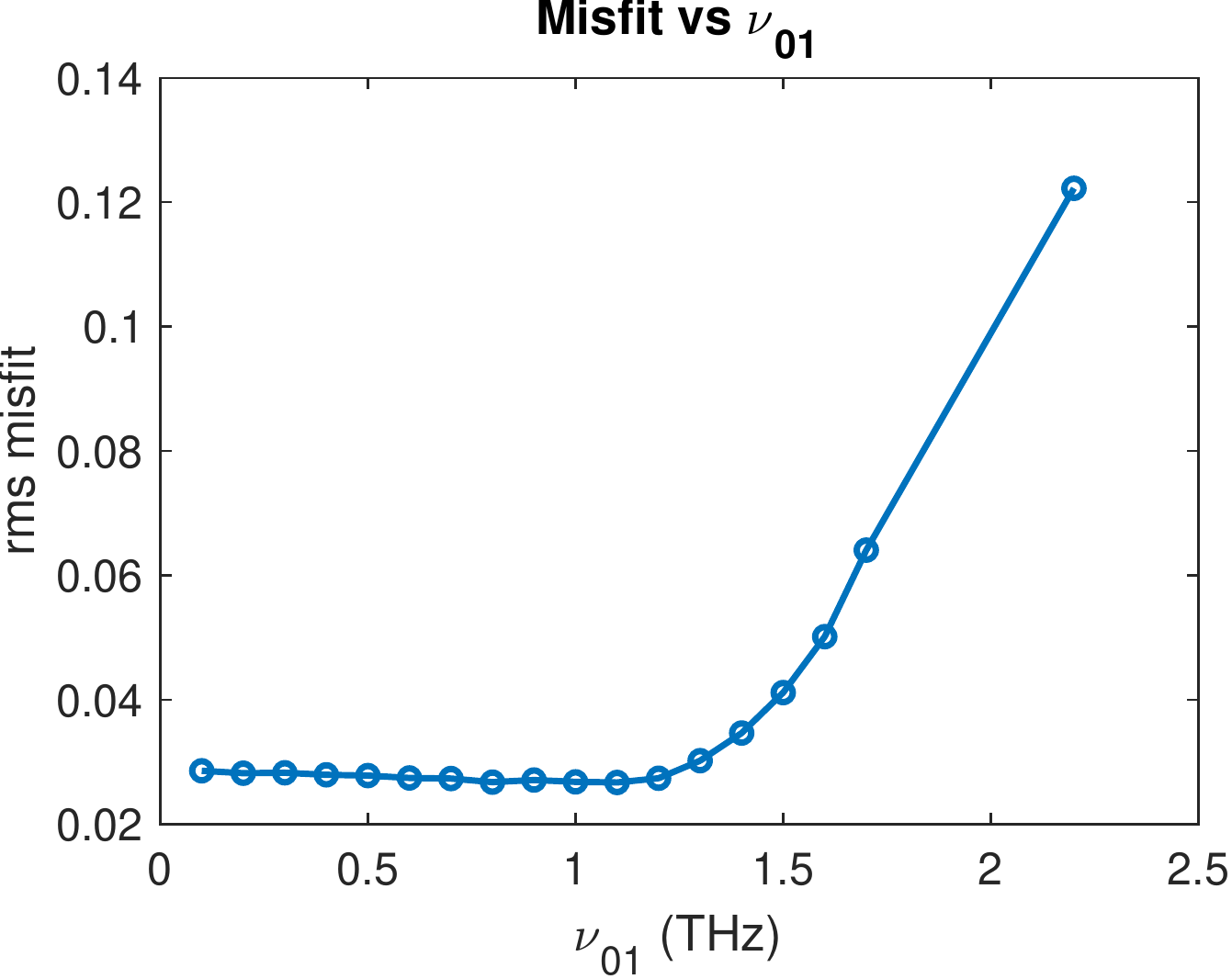}
\caption{Calculated root mean square (rms) misfit between experimental complex reflectivities ($r(\nu) = \sqrt{R(\nu)}\cdot e^{i\varphi(\nu)}$) and fit curves (examples shown in Fig.~\ref{fig:fixnu1spec}) as a function of fixed lowest-mode frequency $\nu_{01}$ (misfit here calculated for the subrange $0.8$-$2\THz$ to concentrate on how this specific spectral range is affected). Here the case of ``omitted $\nu_{01}$'' corresponds to a lowest-mode frequency of $2.2\THz$ (corresponding to $\nu_{02}$ in the other fits).
} \label{fig:fixnu1misfit}
\end{figure}

As discussed in the main paper, the measured strong reflectivity signature at $\nu\simx 1.75\THz$ indicates the presence of a phase mode  (PM) at lower frequency $\nu_{01}$, although we cannot identify its position from our reflectivity spectra.
This is demonstrated in Fig.~\ref{fig:fixnu1spec}, where we show fitted spectra for a set of fixed values for $\nu_{01}$, with the corresponding misfit (calculated here for the subrange 0.8-2~THz) in Fig.~\ref{fig:fixnu1misfit}.
One can see the misfit is essentially flat for $\nu_{01}\lesssim 1.2\THz$, with reasonable fits to the spectra (except in the case where such a mode is omitted).
For this reason, in the main paper, we assume a PM at $\nu_{01}=0.1\THz$, corresponding to the ``pinned phason frequency'' found in previous experimental studies \cite{mihaly1989,degiorgi91}.

\newpage
\subsection{Extended-bandwidth TDS measurements with ABCD detection at $T=20\Kel$}

\begin{figure}[!h]
	\includegraphics[width=1\textwidth]{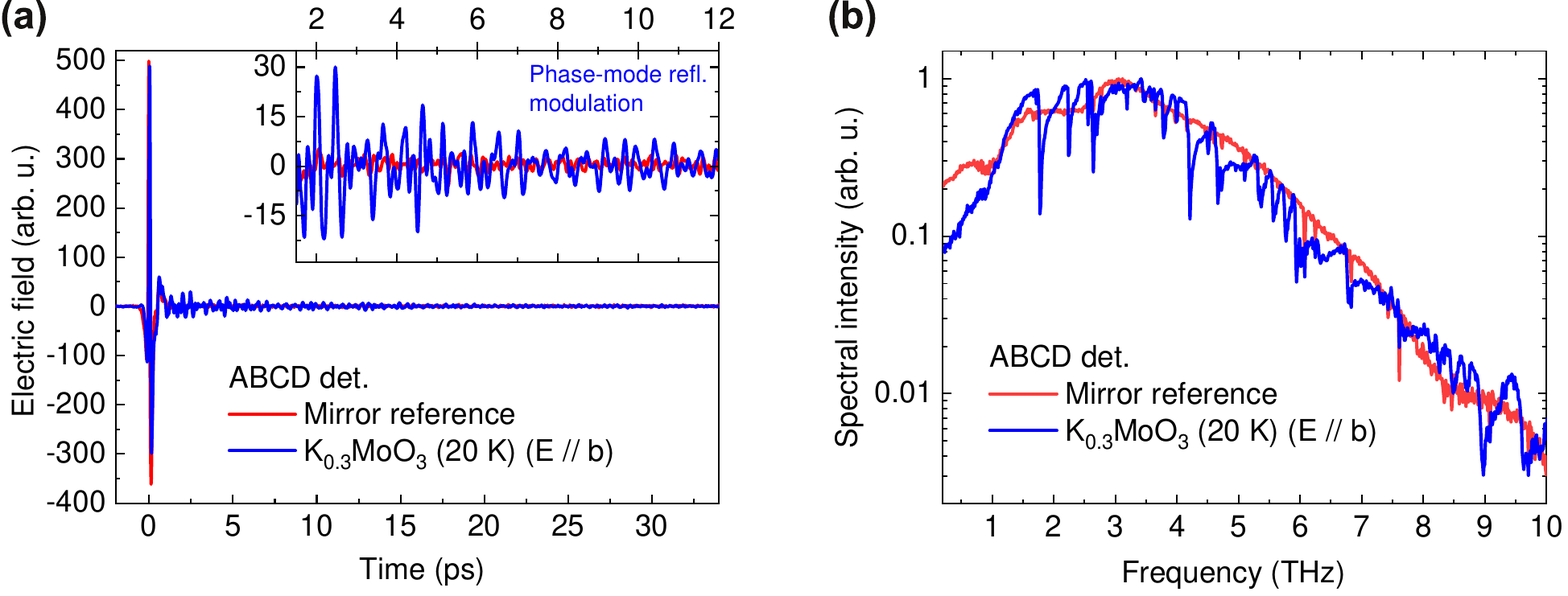}
\caption{(a) Example of detected THz time-domain fields with reflection sample geometry utilizing ABCD: reference (using a mirror in the sample position, red curve), and \kmo sample at $T=20\Kel$ (blue curve). Inset shows magnified range of oscillatory signatures after the main pulse for the \kmo data (while the weak residual oscillations of the mirror reference are due to residual water-vapor absorption in the THz beam path). (b) Corresponding intensity spectra.
} \label{fig:abcdtds}
\end{figure}

The majority of THz-TDS experiments here used EOS detection, providing a detection  bandwidth limited to $\simx 7\THz$ (see Fig.s~2 and 3 in main paper), due to the onset of phonon absorption in the GaP crystal.
To extend the bandwidth and detect higher-lying modes, we also employed the ABCD detection method \cite{karpowicz08}, which does not suffer from such phonon absorption, and allows here detection up to $\simx 10\THz$ (see Fig.~\ref{fig:abcdtds}). 
However, we found that the signal-to-noise ratio and sample alignment sensitivity were superior for EOS detection, and employed ABCD only for $T=20\Kel$ to resolve PMs in the range 7-$9\THz$.

%\newpage

\section{Time-dependent Ginzburg-Landau model}

\subsection{Theoretical description}\label{sec:tdgl_theory}

The theoretical basis of the TDGL model is given in \cite{schaefer10,schaefer14,thomson17} (and their respective Supplementaries), so we summarize the aspects here only briefly to clarify the particular details/notation for the fitted model results in the present paper.

Writing the complex electronic order parameter (EOP) as 
$\tilde\Delta = \Delta e^{i\varphi} = \Delta_1 + i\Delta_2$ 
and complex bare-phonon coordinates
$\tilde\xi_n = \xi_n e^{i\chi_n} = \xi_{n1} + i\xi_{n2}$ ($n=1\dots N$) (where all coordinates refer to the complex wave amplitudes of the $q=2\kF$ components), we consider the potential function:
\beq
U(\tilde\Delta,\tilde\xi_1,\dots,\tilde\xi_N)=
U_{\Delta} + U_{\xi n} + \usr{U}{c} + \usr{U}{p},
\label{eq:tdglU}
\eeq
where 
$U_{\Delta}=-\tfrac{1}{2}\alpha(\Tcz-T)\Delta^2 + \tfrac{1}{4}\beta \Delta^4$ 
is the Mexican hat potential, 
$U_{\xi n}=\tfrac{1}{2}\Wzn^2 \xi_n^2$ 
represents the elastic energy stored in the bare phonon mode $n$ with frequency $\Wzn$, 
$\usr{U}{c}=-m_n(\Delta_1\xi_{n1}+\Delta_2\xi_{n2}) = -m_n\Delta\cdot\xi_{n}\cos(\varphi-\chi_n)$ is the linear coupling term (summations over $n$ are left implicit), and
$\Upin=-\Wpin^2\Delta^2\cos{\varphi}$ 
the impurity pinning energy \cite{tucker1989,wonneberger1999} (discussed below).

This has the equilibrium solution (neglecting a small correction for the pinning potential)
$$\Delta_0^2=\frac{\alpha(\Tc-T)}{\beta}, \qquad 
\xi_{0n}=\frac{m_n}{\Wzn^2}\Delta_0, \qquad 
\Tc=\Tcz+\frac{m_n^2}{\alpha\Wzn^2},
$$
where $\Tc$ is the renormalized critical temperature, and 
$\varphi_0=\chi_{n0}=0$  due to the impurity pinning minimum.

Calculating the Hessian matrix of the potential about $\tilde\Delta_0=\Delta_0$ yields the linearized equations of motion \cite{schaefer10,schaefer14,thomson17}:
\begin{subequations}
\begin{align}
\partial_t{\hat\Delta_1}&=
-\kappa_1 \left[ 2\alpha(\Tc-T) + \frac{m_n^2}{\Wzn^2} +2\Wpin^2 \right]\hat\Delta_1  +\kappa_1 m_n\hat\xi_{n1} \label{eq:motionDA}
%\partial_t^2{\hat\Delta_1}&=-\left[ 2\alpha(\Tc-T) + \frac{m_n^2}{\Wzn^2} \right]\hat\Delta_1 +m_n\hat\xi_{n1} - \gamma_1\partial_t{\hat\Delta_1} \label{eq:motionDA}
 \\
\partial_t^2{\hat\Delta_2}&=-\left(\frac{m_n^2}{\Wzn^2}+\Wpin^2\right)\hat\Delta_2
 +m_n\hat\xi_{n2} - \gamma_2\partial_t{\hat\Delta_2} \label{eq:motionDP}
 \\  %
\partial_t^2{\hat\xi_{n1}}&=m_n\hat\Delta_{1} - \Wzn^2 \hat\xi_{n1} 
- \gamma_{\xi n}\partial_t \hat{\xi}_{n2}\\
\partial_t^2{\hat\xi_{n2}}&=m_n\hat\Delta_{2} - \Wzn^2 \hat\xi_{n2}
- \gamma_{\xi n}\partial_t \hat{\xi}_{n2}
\end{align} \label{eq:motion}%
\end{subequations}
where $\hat\Delta_1\approx \Delta-\Delta_0$ and $\hat\Delta_2\approx\Delta_0 \varphi$ represent the amplitude and phase deviations from equilibrium (likewise for $\hat \xi_{n1}$,$\hat \xi_{n2}$), and we have added phenomenological damping constants $\gamma_{1,2}$ for the EOP $\hat\Delta_{1,2}$ (and allow for $\gamma_1 \neq \gamma_2$), and $\gamma_{\xi n}$ for the $n$th bare phonon.
Note that in Eq.~\eqref{eq:motionDA} we have taken the overdamped limit 
$\partial_t^2\hat\Delta_1 \ll \gamma \partial_t\hat\Delta_1$ 
for the EOP-amplitude (as per \cite{schaefer10,schaefer14,thomson17})
while we retain the general damping case for the EOP-phase in Eq.~\eqref{eq:motionDP}, as appropriate in the present paper where we consider $\gamma_2\ll \gamma_1$.
Also, while tests were performed with different models for the bare-phonon damping $\gamma_{\xi n}> 0$, we found that these do not significantly assist fitting of the experimental modes and set $\gamma_{\xi n}\equiv 0$ for the analysis shown in the main paper.

\begin{figure}[!h]
	\includegraphics[width=1\textwidth]{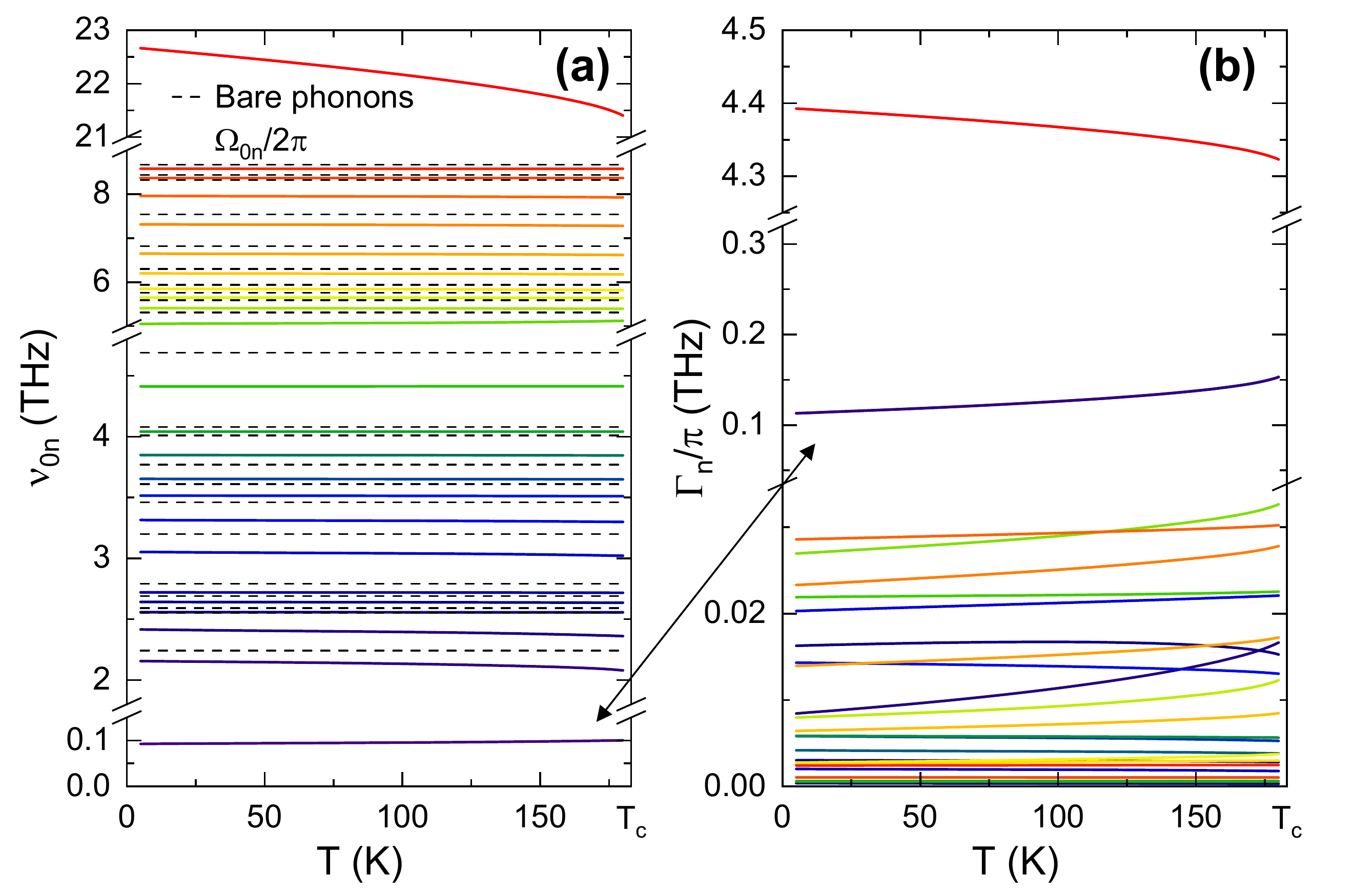}
\caption{PM parameters vs. $T$ from TDGL model of experimental modes: (a) Mode frequencies and (b) FWHM bandwidths.} \label{fig:tdgl_phase}
\end{figure}

The collective modes are found by substituting the ansatz $\propto e^{\lambda t}$ for all coordinates in Eq.s~\eqref{eq:motion}, yielding the eigenvalues $\lambda_n=-\Gamma_n/2+i\wzn$.  
For the amplitude channel, with eigenvector components in $(\hat\Delta_1,\hat\xi_{n1})$, one has $N$ modes with $\wzn>0$ and one overdamped mode ($\wzn=0$).
For the phase channel, with eigenvector components in $(\hat\Delta_2,\hat\xi_{n2})$, one has (i) a ``phason'' with $\omega_{01}$ close to zero 
(at finite frequency due to pinning when the damping $\gamma_2$ is sufficiently weak, as is the case here), (ii) $N-1$ modes close to their respective bare-phonon frequencies ($\Wzn$, $n \geq 2$) and (iii) one high-frequency mode (well above all $\Wzn$) with a much higher damping.
(All modes with $\wzn\neq 0$ also possessing a complex-conjugate eigenvalue with $\wzn<0$).
These results are depicted in Fig.~\ref{fig:tdgl_phase} for the TDGL parameters employed in the main paper to model the experimental data. %

An inspection of the eigenvector for the high-frequency PM above 20~THz shows that it involves almost purely the EOP phase ($\hat\Delta_2$), and only manifests here by not taking the overdamped limit for the phase channel in Eq.~\eqref{eq:motion}.
Given its energy, one might be tempted to connect it with the single-particle gap (also found close to this frequency \cite{degiorgi94}), although we stress here that the TDGL does not explicitly contain this gap energy, and so one cannot make this association. 
Indeed, the condensation energy in Eq.~\eqref{eq:tdglU} can be shown to be 
$U_C=\tfrac{1}{4}\alpha(T-T_c)\Delta_0^2$, where the arbitrary nominal scaling of $\Delta$ reflects this lack of energy-gap calibration. 
Given that this predicted mode has a frequency close to the single-particle gap, it may well be more strongly damped than in the TDGL prediction.  Also, its current position is based on a TDGL model accounting for bare modes only up to $\simx 9\THz$, and so such a mode may rather appear at even higher frequencies if additional bare modes were included.
Interestingly, a close inspection of this spectral region in previous 
mid-infrared reflectivity studies \cite{degiorgi91,degiorgi94,beyer2012} indicates that such a feature could be possibly present, but obscured due to overlap with other spectral features.

The ansatz for our pinning potential $\Upin$ here warrants further discussion.  An intuitive formulation would be 
$\Upinx=-\usr{V}{i}\cdot\Delta\cdot\cos{\varphi}=\usr{V}{i}\cdot\Delta_1$ \cite{tucker1989}, with the factor $\Delta$ reflecting the CDW amplitude (and hence the magnitude of net charge interacting with the impurity) and $\cos{\varphi}$ describing how this interaction varies between attractive/repulsive as the CDW translates over the impurity.  Clearly, for the form $\Upinx$ (with $\usr{V}{i}$ constant), the Hessian terms $\partial^2 \Upinx/\partial \Delta_{1,2}^2$ vanish, and hence also the force terms about the equilibrium $\Delta_0$ in Eq.s~\eqref{eq:motion} (the pinning only causing a small positive shift in $\Delta_0$).
A more careful treatment of the relation between the phase $\varphi$ and Cartesian coordinate $\hat \Delta_2$ about $(\Delta_1,\Delta_2)=(\Delta_0,0)$ yields 
$\Upinx\approx -\usr{V}{i}\cdot\Delta_0(1-\tfrac{1}{2}\varphi^2)
\approx -\usr{V}{i}(\Delta_0-\tfrac{1}{2}\Delta_2^2/\Delta_0)$.
This indeed yields a finite value of
$\partial^2 \Upinx/\partial \Delta_2^2=-\usr{V}{i}/\Delta_0$.  However, this term diverges as $T\rarr T_c$ ($\Delta_0\rarr 0$), which for \kmo seems untenable as the experimental phason frequency is found to remain close to 
$\nuzP=0.1\THz$ for all $T$ \cite{degiorgi91} (although divergent pinning fields for $T\rarr T_c$ were indeed observed for other quasi-1D CDW system, e.g. \chem{Nb/,Se/3} \cite{mccarten1992}).

We instead propose a pinning potential of the form 
$\Upin=-\Wpin^2\Delta^2\cos{\varphi}=-\Wpin^2\Delta\cdot\Delta_1$.
This yields 
\beq
\frac{\partial^2\Upin}{\partial \Delta_1^2}=-\Wpin^2
\frac{\Delta_1(2\Delta_1^2+3\Delta_2^2)}{\Delta^2}
\xrightarrow[\Delta_0]{} - 2\Wpin^2, \qquad 
\frac{\partial^2\Upin}{\partial \Delta_2^2}=-\Wpin^2\frac{\Delta_1^3}{2\Delta_1^2+3\Delta_2^2}
\xrightarrow[\Delta_0]{} - \Wpin^2,
\eeq
and hence a $T$-independent pinning force along both $\Delta_{1,2}$.  Note that for our TDGL parameters, the pinning term only has a significant effect on the lowest PM (phason).

\newpage
\subsection{Fitted TDGL parameters}

%In Table~\ref{tab:tdglpars} we list the TDGL parameters used for fitting the experimental AM and PM.

\begin{table}[htbp]
% \def \colwidA{0.1\columnwidth}
% \def \colwidB{0.06\columnwidth}
    % \centering
    \begin{center}
        % \begin{tabular}{|c|c|c|c|c|c|c|c|}
        \begin{tabular}{|C{0.06\textwidth}|C{0.15\textwidth}|C{0.15\textwidth}|C{0.08\textwidth}|C{0.12\textwidth}|C{0.08\textwidth}|C{0.08\textwidth}|c|}
            \hline
            $n$ & $\Wzn$ & $\kappa_1 m_n^2$ & $b_n$ & $\alpha$ & $\gamma_1/2\pi$ & $\gamma_2/\gamma_1$ & $\Wpin$ \\
             &  (THz) & (THz$^3$) & & (ps$^{-2}$~K$^{-1}$) & (THz) & & (THz)\\
            \hline
            1 & 1.82 (1.79) & 1170 (580) & 0.30 (0) & \multirow{23}{*}{\shortstack{72.5 \\ (46)}} & \multirow{23}{*}{\shortstack{52.5 \\ (46.8)}} & \multirow{23}{*}{\shortstack{0.09 \\ (1)}} & \multirow{23}{*}{\shortstack{0.7 \\ (9.3)}}\\
            2 & 2.24 (2.25) & 240 (320) & 0 (0) & & & & \\
            3 & 2.55 (2.64) & 60 (1150) & 0 (0) & & & & \\
            4 & 2.59 & 250 & \multirow{10}{*}{0} & & & & \\
            5 & 2.69 & 160 &  & & & & \\
            6 & 2.79 & 610 &  & & & & \\
            7 & 3.20 & 410 &  & & & & \\
            8 & 3.46 & 360 &  & & & & \\
            9 & 3.61 & 340 &  & & & & \\
            10 & 3.77 & 800 &  & & & & \\
            11 & 4.01 & 1200 &  & & & & \\
            12 & 4.08 & 1700 &  & & & & \\
            13 & 4.69 & 3300 &  & & & & \\
            14 & 5.31 & 1880 & 0.35 & & & & \\
            15 & 5.59 & 2280 & 0.30 & & & & \\
            16 & 5.76 & 3750 & 0.30 & & & & \\
            17 & 5.94 & 5250 & \multirow{7}{*}{0} & & & & \\
            18 & 6.30 & 4050 &  & & & & \\
            19 & 6.82 & 6200 &  & & & & \\
            20 & 7.54 & 7350 &  & & & & \\
            21 & 8.32 & 8500 &  & & & & \\
            22 & 8.44 & 12000 &  & & & & \\
            23 & 8.67 & 15700 &  & & & & \\
         \hline
    \end{tabular}
    \end{center}

    \caption{Overview of TDGL model parameters used for obtaining the $T$-dependent fit curves in Fig. 5 in the main paper, compared to the values used in our previous work (in brackets) \cite{thomson17} (with $\kappa_1 = \gamma_1^{-1}$ and other parameters as defined above in Sec.~\ref{sec:tdgl_theory} and in the main paper). Additionally, the $T$-dependent parameters of both EOP-phase damping and impurity pinning potential (defined in \cite{thomson17}) were both set to zero in this work, as these did not improve the quality of fitting PMs vs $T$.}
    \label{tab:tdglpars}
\end{table}

%\bibliography{references}
%\bibliography{CPA_Lab}